\documentclass[preprint]{aastex}

\usepackage{epsf}

\def\linebreak{\hfil\break}

\def\
{\hfil\linebreak}

\def\degree{\ifmmode {^\circ}\else {$^\circ$}\fi}
\def\mum{\ifmmode {\rm \mu {\rm m}}\else $\rm \mu {\rm m}$\fi}
\def\arcsec{\ifmmode ^{\prime \prime}\else $^{\prime \prime}$\fi}

\def\inch{\ifmmode ^{\prime \prime}\else $^{\prime \prime}$\fi}
\def\arcmin{\ifmmode ^{\prime}\else $^{\prime}$\fi}

\def\mjup{\ifmmode {\rm M_J}\else $\rm M_J$\fi}
\def\rjup{\ifmmode {\rm R_J}\else $\rm R_J$\fi}
\def\mearth{\ifmmode {\rm M_{\oplus}}\else $\rm M_{\oplus}$\fi}
\def\rearth{\ifmmode {\rm R_{\oplus}}\else $\rm R_{\oplus}$\fi}
\def\lsun{\ifmmode {\rm L_{\odot}}\else $\rm L_{\odot}$\fi}
\def\msun{\ifmmode {\rm M_{\odot}}\else $\rm M_{\odot}$\fi}
\def\mjupyr{\ifmmode {\rm M_J~yr^{-1}}\else $\rm M_J~yr^{-1}$\fi}
\def\msunyr{\ifmmode {\rm M_{\odot}~yr^{-1}}\else $\rm M_{\odot}~yr^{-1}$\fi}
\def\kms{\ifmmode {\rm km~s^{-1}}\else $\rm km~s^{-1}$\fi}

\def\mstar{$M_\star$}

\def\qdstar{$Q_d^\star$}

\def\gcm3{g~cm$^{-3}$}

\def\powlaw{{\rm n}} 
\def\Phid{\Phi_d}
\def\rin{{R_{\rm in}}}
\def\rout{{R_{\rm out}}}

\newbox\grsign \setbox\grsign=\hbox{$>$} \newdimen\grdimen \grdimen=\ht\grsign
\newbox\simlessbox \newbox\simgreatbox
\setbox\simgreatbox=\hbox{\raise.5ex\hbox{$>$}\llap
     {\lower.5ex\hbox{$\sim$}}}\ht1=\grdimen\dp1=0pt
\setbox\simlessbox=\hbox{\raise.5ex\hbox{$<$}\llap
     {\lower.5ex\hbox{$\sim$}}}\ht2=\grdimen\dp2=0pt

\def\simless{\mathrel{\copy\simlessbox}}

\begin{document}

\title{A New Hybrid N-Body-Coagulation Code for the Formation of Gas Giant Planets}
\vskip 7ex
\author{Benjamin C. Bromley}
\affil{Department of Physics, University of Utah, 
201 JFB, Salt Lake City, UT 84112} 
\email{e-mail: bromley@physics.utah.edu}
\author{Scott J. Kenyon}
\affil{Smithsonian Astrophysical Observatory,
60 Garden Street, Cambridge, MA 02138} 
\email{e-mail: skenyon@cfa.harvard.edu}
%
%

\begin{abstract}

We describe an updated version of our hybrid $N$-body-coagulation code for planet formation.
In addition to the features of our 2006--2008 code, our treatment now includes algorithms
for the 1D evolution of the viscous disk, the accretion of small particles in planetary
atmospheres, gas accretion onto massive cores, and the response of $N$-bodies to the 
gravitational potential of the gaseous disk and the swarm of planetesimals.  To validate the 
N-body portion of the algorithm, we use a battery of tests in planetary dynamics. As a first
application of the complete code, we consider the evolution of Pluto-mass planetesimals in 
a swarm of 0.1--1~cm 
pebbles. In a typical evolution time of 1--3~Myr, our calculations transform 0.01--0.1 \msun\ disks 
of gas and dust into planetary systems containing super-Earths, Saturns, and Jupiters.  Low 
mass planets form more often than massive planets; disks with smaller $\alpha$ form more 
massive planets than disks with larger $\alpha$.  For Jupiter-mass planets, masses of solid 
cores are 10--100~\mearth. 

\end{abstract}

\subjectheadings{planetary systems -- solar system: formation -- 
stars: formation -- circumstellar matter}

\section{INTRODUCTION}

Gas giant planets form in gaseous disks surrounding young stars. In the
`core accretion' theory, collisions and mergers of solid planetesimals
produce 1--10 \mearth\ (Earth mass) icy cores which rapidly accrete gas 
\citep[e.g.,][]{pol96,ali05,ida05,cha07,lis07,dodson2008,dang2010}.
As the planets grow, viscous transport and photoevaporation also remove 
material from the disk \citep[e.g.,][]{alex2009, owen2010}.  Eventually, these
processes exhaust the disk, leaving behind a young planetary system.

Observations place many constraints on this theory. Although nearly all of the 
youngest stars are surrounded by massive disks of gas and dust, very few stars 
with ages of 3--10~Myr have dusty disks \citep[e.g.,][]{hai01,cur09,kenn09,mamajek2009}.
Thus, gas giant planets must form in 1--3 Myr. Stars with ages of $\sim$ 1~Myr 
have typical disk masses of 0.001--0.1 \msun\ and disk radii of 10--200 AU
\citep[e.g.,][]{and05,ise09}, setting the local environment where planets form.
For stars with ages of $\sim$ 1 Gyr, the frequency of ice/gas giant planets
around low mass stars is $\sim$ 20\%--40\% \citep{cum08,gou2010}.  Although 
many known exoplanets are gas giants with masses comparable to Jupiter, lower 
mass planets outnumber Jupiters by factors of $\sim$ 10 
\citep{may09,bor10,gou2010,hol2010,how2010}. 

Despite the complexity of the core accretion theory, clusters of computers are 
now capable of completing an end-to-end simulation of planet formation in an 
evolving gaseous disk. Our goal is to build this simulation and to test the 
core accretion theory. Over the past decade, we have developed a hybrid, multiannulus 
$N$-body--coagulation code for planet formation \citep{kb04a,bk06,kb08}. The 
centerpieces of our approach are a multiannulus coagulation code -- which treats
the growth and dynamical evolution of small objects in 2D using a set of statistical 
algorithms -- and an $N$-body code -- which follows the 3D trajectories of massive 
objects orbiting a central star. With additional software to combine the two codes
into a unified whole, we have considered the formation of terrestrial planets 
\citep{kb06} and Kuiper belt objects \citep{kb04c,kbod08} around the Sun and the formation 
and evolution of debris disks around other stars \citep[][2004b, 2008, 2010]{kb04a}.

Treating the formation of gas giant planets with our 2008 code \citep{kb08} requires 
algorithms for additional physical processes. As described in \S2, we added 
(i) a 1D solution for the evolution of a viscous disk on a grid extending from 0.2~AU
to $10^4$~AU \citep{cha2009,alex2009},
(ii) algorithms for accretion of small particles encountering
a planetary atmosphere \citep{ina2003} and for accretion of gas
onto massive cores, 
(iii) a prescription for the evolution of the radius and luminosity
of a gas giant planet accreting material from a disk, and
(iv) new code to treat the gravitational potential of the gaseous disk
and the swarm of planetesimals in the $N$-body code.
Tests of these additions demonstrate that our complete code reproduces 
the simulations of viscous disks in \citet{alex2009}, 
accretion rates for small particles in \citet{ina2003} and for gas in \citet{dang2010},
and the battery of $N$-body code tests from \citet{bk06}. In \S3, we show that
our code also reproduces results for planets migrating through planetesimal disks 
\citep{hah99,kir09}. 

As a first application to gas giant planet formation, we consider whether ensembles of
Pluto-mass cores embedded in gaseous disks with masses of 0.01--0.1 \msun\ can 
become gas giant planets. Our results in \S4 demonstrate that a disk of Pluto-mass 
objects will not form gas giants in 10~Myr. However, a few Pluto-mass seeds embedded 
in a disk of 0.1--1~cm pebbles can evolve into a planetary system with super-Earths, 
Saturns, and Jupiters. If we neglect orbital migration, ice and gas giants form at 
semimajor axes of 1--100 AU on timescales of 1--3 Myr. Thus, these calculations match 
some of the observed properties of exoplanets without violating the constraints imposed 
by observations of gaseous disks around the youngest stars.

\section{THE MODEL}

\subsection{Disk evolution}\label{subsect:diskevolve}

Disk evolution sets the context for our planet formation calculations. 
Viscous processes within the disk transport mass inwards onto the 
central star and angular momentum outwards into the surrounding
molecular cloud. Heating from the central star modifies the
temperature structure within the disk and removes material from the
upper layers of the disk. All of these processes change disk structure
on timescales comparable to the growth time for planetesimals.  Thus,
planets form in a rapidly evolving disk.

For a disk with surface density $\Sigma$ and viscosity $\nu$,
conservation of angular momentum and energy leads to a non-linear
diffusion equation for the time evolution of $\Sigma$
\citep[e.g.,][]{lbp1974,pri1981},
\begin{equation}
\frac{\partial \Sigma}{\partial t} = 3 R^{-1}~\frac{\partial}{\partial R} ~ \left ( R^{1/2}~\frac{\partial}{\partial R} ~ \{ \nu \Sigma R^{1/2} \} \right ) - \left ( \frac{\partial \Sigma}{\partial t} \right )_{ext} ~ ,
\label{eq: disk-evol}
\end{equation}
where $R$ is the radial distance from the central star and $t$ is the time.
The first term is the change in $\Sigma$ from viscous evolution; the
second term is the change in $\Sigma$ from other processes such as
photoevaporation \citep[e.g.,][]{alex2006, owen2010} or planet formation
\citep[e.g.,][]{alex2009}.  The viscosity is $\nu = \alpha c_s H$,
where $c_s$ is the sound speed, $H$ is the vertical scale height of
the disk, and $\alpha$ is the viscosity parameter.  The sound speed is
$c_s^2 = \gamma k T_d / \mu m_H$, where $\gamma$ is the ratio of
specific heats, $k$ is Boltzman's constant, $T_d$ is the midplane
temperature of the disk, $\mu$ is the mean molecular weight, and $m_H$
is the mass of a hydrogen atom. The scale height of the disk is $H$ =
$c_s \Omega^{-1}$, where $\Omega = G M_{\star}/R^3$ is the angular
velocity.

There are two approaches to solving equation (\ref{eq:
  disk-evol}). Analytic solutions adopt a constant mass flow rate
$\dot{M} = 3 \pi \nu \Sigma$ through the disk, approximate a vertical
structure, and solve directly for $\Sigma (R,t)$, $T_d(R,t)$, and
other physical variables. Numerical solutions either adopt or solve
for the vertical structure and then solve for the time variation of
$\dot{M}$ and other physical variables \citep[e.g.,][]{hues2005,mitch10}.

Here, we consider planet formation using both approaches. For the
analytic solution, we follow \citet{cha2009}, who derives an elegant
time dependent model for a viscous disk irradiated by a central
star. For the numerical solution, we assume that the midplane temperature 
is derived from the sum of the energy generated by viscous dissipation
(subscript "V") and the energy from irradiation (subscripit "I")
\begin{equation}
T_d^4 = T_V^4 + T_I^4 ~ .
\label{eq: tdisk}
\end{equation}
The viscous temperature is
\begin{equation}
T_V^4 = \frac{27 \kappa \nu \Sigma^2 \Omega^2}{64 \sigma} ~ ,
\label{eq: t-visc1}
\end{equation}
where $\kappa$ is the opacity and $\sigma$ is the Stefan-Boltzmann constant
\citep[e.g.,][]{ruden1986,ruden1991}. With $\nu = \alpha c_s^2 \Omega^{-1}$ and 
$t_2 = (27 \alpha /64 \sigma) ~ (\gamma k / \mu m_H) ~ \kappa \Omega \Sigma^2$,
the viscous temperature is
\begin{equation}
T_V^4 = t_2 T_d.
\label{eq: t-visc2}
\end{equation}

If the disk is vertically isothermal, the irradiation temperature is
$T_I^4(R) = (\theta / 2) (R_{\star} / R)^2 T_{\star}$, where $R_{\star}$ and
$T_{\star}$ are the radius and effective temperature of the central star and
\begin{equation}
\theta = \frac{4}{3 \pi} \left ( \frac{R_{\star}}{R} \right )^3 + R ~ \frac{\partial (H/R)}{\partial R} 
\end{equation}
\citep{chiang1997}. Thus, the irradiation temperature is 
\begin{equation}
\left ( \frac{T_I}{T_\star} \right)^4 = \frac{2}{3 \pi} \left ( \frac{R_{\star}}{R} \right )^3 + \frac{H}{2R} \left ( \frac{R_{\star}}{R} \right )^2 \left ( \frac{\partial~{\rm ln}~H}{\partial~{\rm ln}~R} - 1 \right ) ~ .
\label{eq: t-irrad1}
\end{equation}
Following \citet{chiang1997} and \citet{hues2005}, we set
$\partial~{\rm ln}~H / \partial~{\rm ln}~R$ = 9/7.  With $H = c_s
\Omega^{-1}$, we set $t_0 = (2 T_\star^4/3 \pi) (R_{\star}/R)^3$ and
$t_1 = (R_{\star}/R)^2 ~ (7 R \Omega)^{-1} ~ (\gamma k/\mu m_H)^{1/2}
~ T_{\star}^4$.  The irradiation temperature is then
\begin{equation}
T_I^4 = t_0 + t_1 T_d^{1/2} ~ .
\label{eq: t-irrad2}
\end{equation}
Although viscous disks are not vertically isothermal
\citep{ruden1991,dale1998}, this approach yields a reasonable
approximation to the actual disk structure.

Because $T_V$ and $T_I$ are functions of the midplane temperature, we
solve equation (\ref{eq: tdisk}) with a Newton-Raphson technique.
Using equations (\ref{eq: t-visc2}) and (\ref{eq: t-irrad2}), we
re-write equation (\ref{eq: tdisk}) as
\begin{equation}
f(T_d) = T_d^4 - (t_0 + t_1 T_d^{1/2} + t_2 T_d) = 0 ~ .
\label{eq: tdisk-nr}
\end{equation}
Adopting an initial $T_d \approx t_2^{1/3}$ or $T_d \approx
t_1^{2/7}$, the derivative
\begin{equation}
\frac{\partial f}{\partial T_d} = 4 T_d^3 - \frac{t_1}{2} T_d^{-1/2} - t_2
\label{eq: deriv-nr}
\end{equation}
allows us to compute 
\begin{equation}
\delta T_d = f \left ( \frac{\partial f}{\partial T_d} \right )^{-1} ~
\label{eq: delta-t}
\end{equation}
and yields a converged $T_d$ to a part in $10^8$ in 2--3 iterations.

In the inner disk, the temperature is often hot enough to vaporize
dust grains. To account for the change in opacity, we follow
\citet{cha2009} and assume an opacity
\begin{equation}
\kappa = \kappa_0 \left ( \frac{T_d}{T_e} \right )^n
\label{eq: kappa-evap}
\end{equation}
with $n = -14$ in regions with $T_d > T_e =$ 1380 K
\citep{ruden1991,step1998}. For simplicity, we assume $\kappa =
\kappa_0$ when $T_d < T_e$.

To solve for the time evolution of $\Sigma$, we use an explicit
technique with $N$ annuli on a grid extending from $x_{in}$ to
$x_{out}$ where $x$ = 2 $R^{1/2}$ \citep[][1982]{bath1981}.  We adopt
an initial surface density
\begin{equation}
\Sigma_0 = \frac{M_{d,0}}{2 \pi R R_0} e^{-R/R_0} ~ ,
\label{eq: sigma0}
\end{equation}
where $M_{d,0}$ is the initial disk mass and $R_0$ is the initial disk
radius \citep{hart1998,alex2009}.  For the thermodynamic variables, we
adopt $\gamma$ = 7/5 and $\mu$ = 2.4.

\begin{figure}
\includegraphics[width=6.5in]{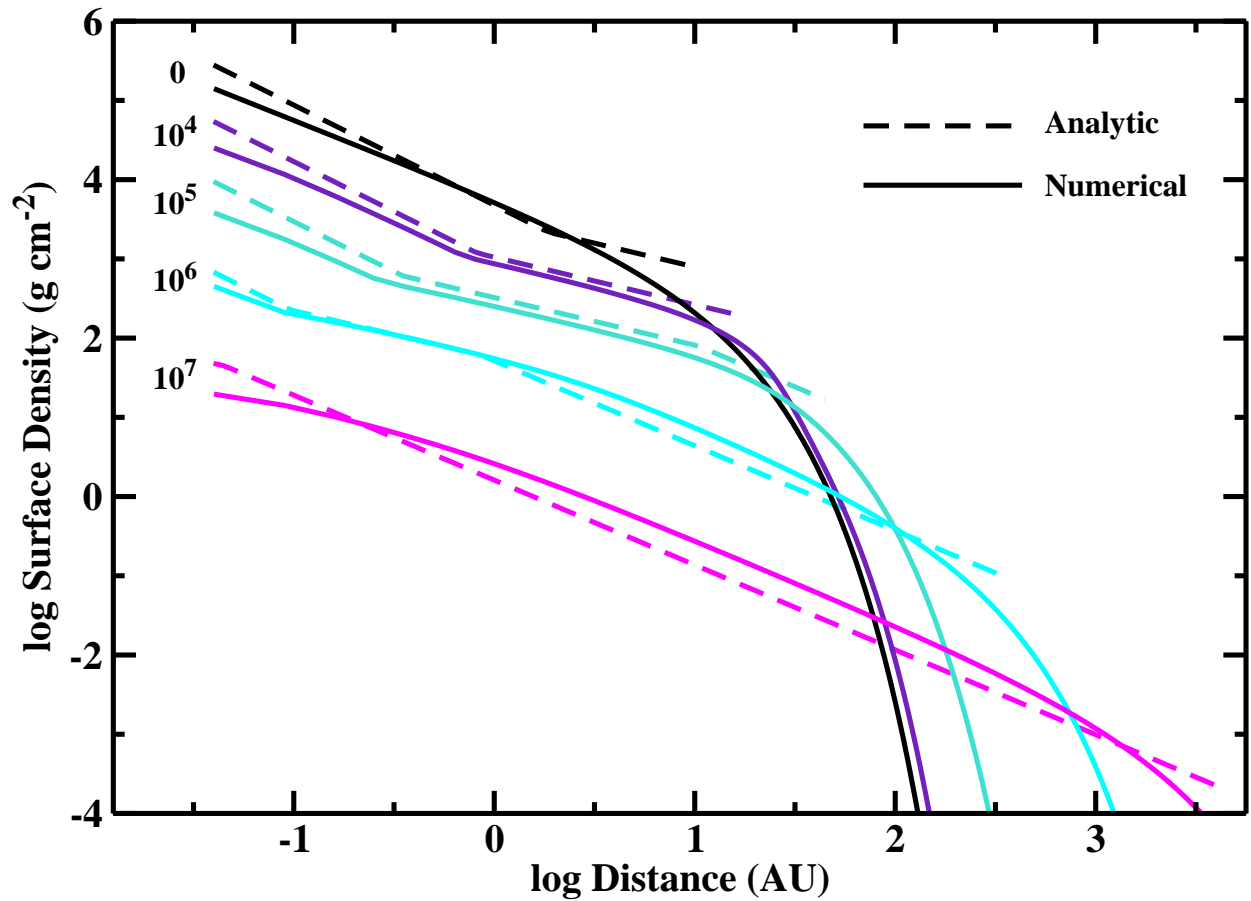}
\vskip 3ex
\caption{%
Time evolution of the surface density of a gaseous disk 
($M_{d,0}$ = 0.04~\msun, $R_{d,0}$ = 10~AU, $\alpha$ = $10^{-2}$)
surrounding a 1 \msun\ star.  Dashed lines show results for the analytic 
disk model of \citet{cha2009}; solid lines show results for our numerical 
solution of the diffusion equation. Despite small differences in the initial 
conditions, the numerical solution tracks the analytic model.
\label{fig: disk1}
}
\end{figure}

Figure \ref{fig: disk1} compares analytic and numerical results for a
disk with $\alpha = 10^{-2}$, $M_{d,0}$ = 0.04 \msun, and $R_0$ =
10~AU surrounding a star with \mstar\ = 1 \msun. The numerical
solution tracks the analytic model well.  At early times, the surface
density declines steeply in the inner disk (where dust grains
evaporate) and more slowly in the outer disk (where viscous transport
dominates).  At late times, irradiation dominates the energy budget;
the surface density then falls more steeply with radius.

\begin{figure}
\includegraphics[width=6.5in]{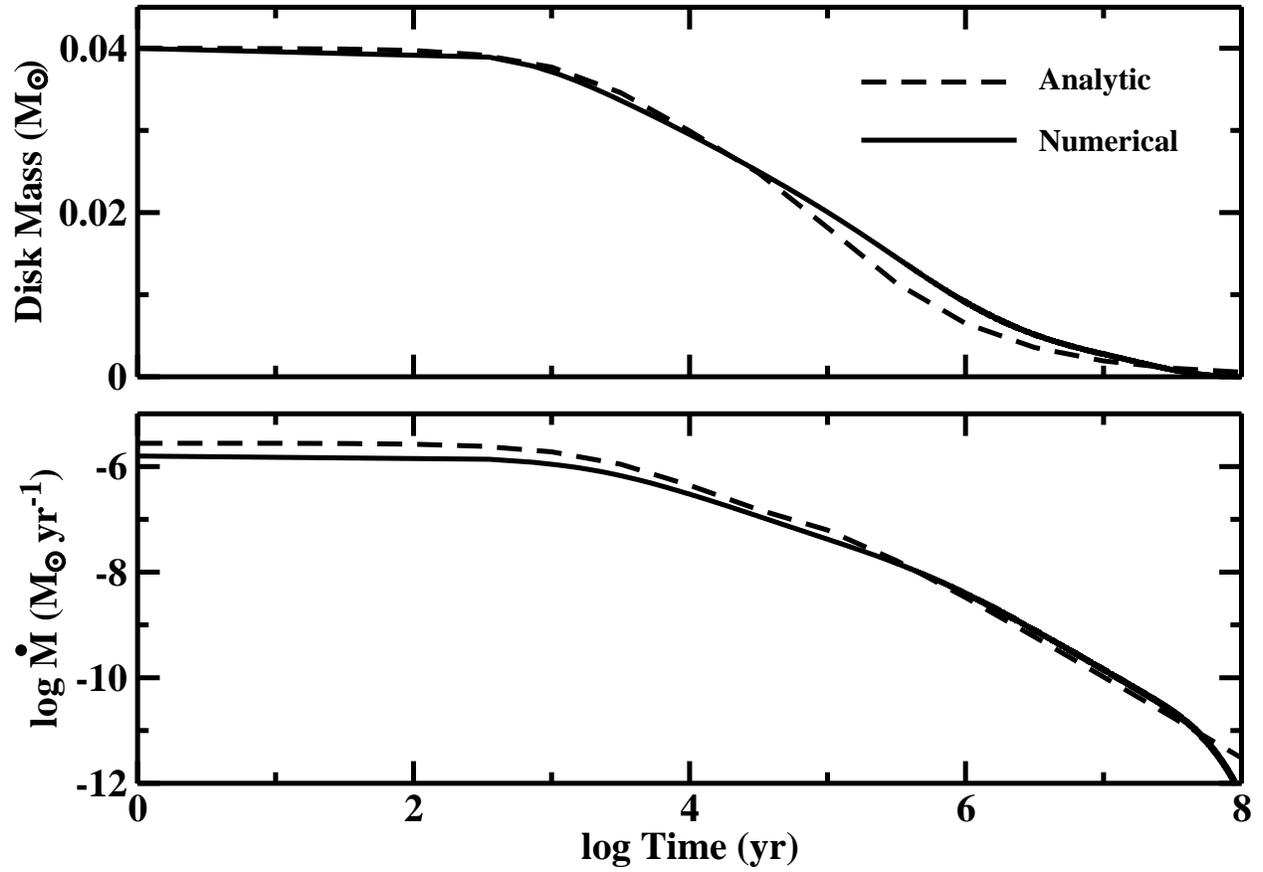}
\vskip 3ex
\caption{%
Time evolution of the disk mass (upper panel) and disk accretion rate onto
the central star (lower panel) for the analytic and numerical solutions in
Figure 1.
\label{fig: disk2}
}
\end{figure}

Figure \ref{fig: disk2} compares the evolution of the disk mass and
accretion rate at the inner edge of the disk. In both solutions, the
disk mass declines by a factor of roughly two in 0.1 Myr, a factor of
roughly four in 1 Myr, and a factor of roughly ten in 10 Myr. Over the
same period, the mass accretion rate onto the central star declines by
roughly four orders of magnitude.

In addition to viscous evolution, we consider photoevaporation of disk
material.  Following \citet{alex2007}, we assume high energy photons
from the central star ionize the disk and drive a wind. As long as the
inner disk is optically thick, recombination powers a `diffuse wind,'
where the change in surface density is
\begin{equation}
(\frac{\partial \Sigma}{\partial t})_{diffuse} = -2 n_0(R) u_0(R) \mu m_H ~ .
\label{eq: diffuse}
\end{equation}
In this expression, 
\begin{equation}
n_0(R) = 0.14 \left ( \frac{3 \Phi_\star}{4 \pi \alpha_B R_g^3} \right )^{1/2} \left ( \frac{2}{(R/R_g)^{15/2} + (R/R_g)^{25/2}} \right )^{1/5} ~ ,
\label{eq: density}
\end{equation}
and 
\begin{equation}
u_0(R) = 0.3423 ~ c_{s,w} ~ \left ( \frac{R}{R_g} - 0.1 \right )^{0.2457} ~ e^{-0.3612~(R/R_g - 0.1)} ~ .
\label{eq: velocity}
\end{equation}
The recombination coefficient is $\alpha_B = 2.58 \times 10^{-13}$
cm$^3$ s$^{-1}$ \citep{cox2000}.  The gravitational radius is $R_g = G
M_\star / c_{s,w}^2$, where $c_{s,w}$ = 10 \kms\ is the sound speed in
the wind. For typical stars, the luminosity of H-ionizing photons,
$\Phi_{\star}$, is $\sim$ $10^{40}$--$10^{42}$ s$^{-1}$. However,
\citet{owen2010} show that X-rays drive a more powerful wind than
lower energy photons. Here, we assume that the mass loss rate in the
wind, $\dot{M}_{wind} = \int 2 \pi R ~ \partial \Sigma / \partial t ~
dR $, is a free parameter and consider disk evolution for a range in
$\dot{M}_{wind}$. This approach is equivalent to assuming a range in
$\Phi_\star$.

As the disk evolves, the diffuse wind removes more and more material
throughout the disk. Eventually, the inner disk becomes optically thin
to ionizing photons. Ionization then drives a `direct wind.' The change 
in surface density with time is then much larger,
\begin{equation}
(\frac{\partial \Sigma}{\partial t})_{direct} = 0.47 ~ \mu ~ m_H ~ c_{s,w} \left ( \frac{\Phi_\star}{4 \pi \alpha_B (H/R) R_{in}^3(t)} \right )^{1/2} \left ( \frac{R}{R_{in}(t)} \right )^{-2.42} ~~,~~ R > R_{in},
\label{eq: direct}
\end{equation}
where $R_{in}(t)$ is the inner edge of the optically thick portion of
the disk. Inside $R_{in}(t)$, the mass loss rate from the disk is
zero.

For mass loss rates $\dot{M}_{wind} \approx 10^{-10} - 10^{-8}$ \msunyr, the 
surface density evolution of a viscous disk with photoevaporation follows 
the evolution in Figure \ref{fig: disk1} for $\sim 10^5 - 10^6$ yr. As the
accretion rate through the disk drops, mass loss from the wind becomes more
and more important. Once the inner disk becomes optically thin, the direct 
wind rapidly empties the inner disk of material. As the system evolves, the 
size of this inner `hole' grows from $R_h \approx R_g \approx$ 3~AU to 
$R_h \approx$ 30--100~AU in 0.01--0.1 Myr \citep[see also][]{alex2009,owen2010}. 
A few thousand years later, the gaseous disk is gone.

\begin{figure}
\includegraphics[width=6.5in]{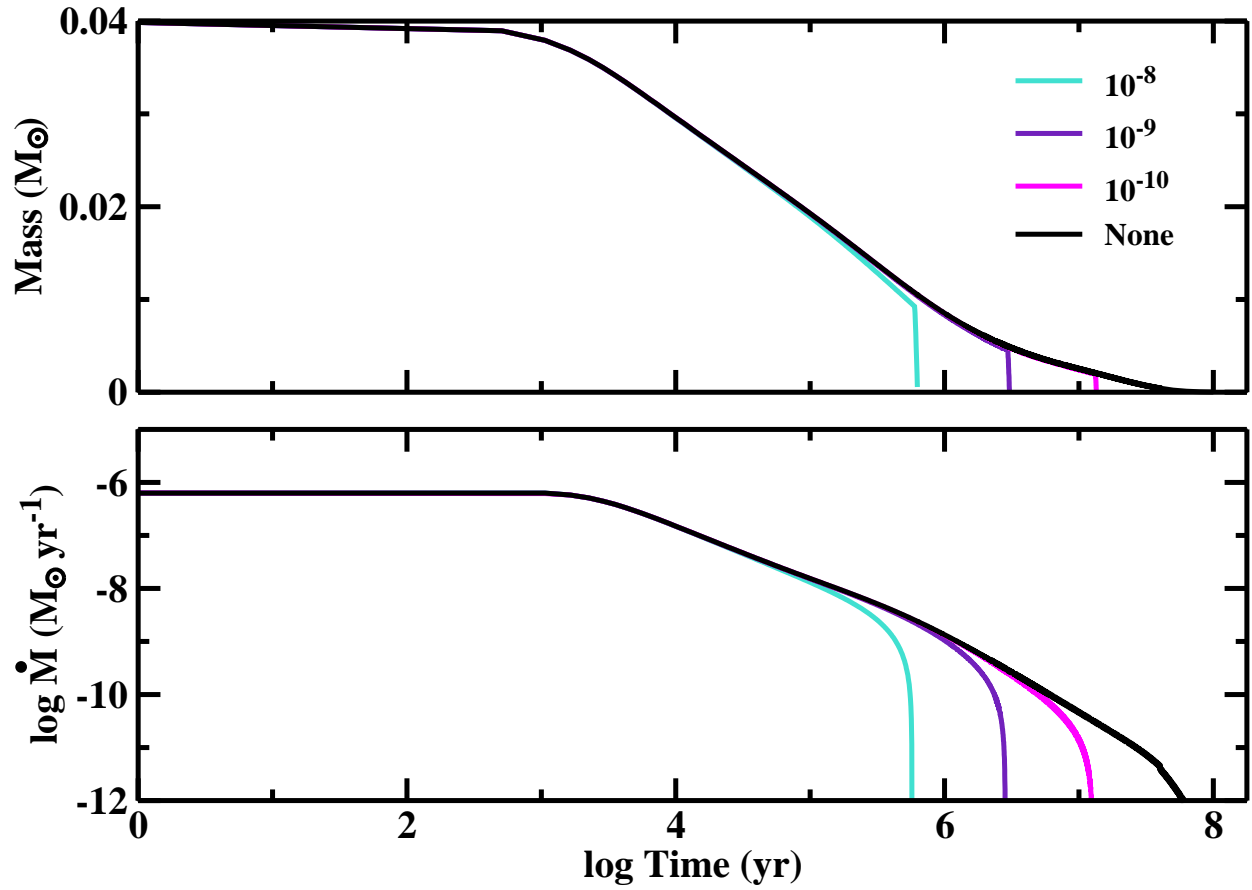}
\vskip 3ex
\caption{%
As in Figure \ref{fig: disk2} for disks with photoevaporative winds. The
legend in the upper panel indicates the mass loss rate of the wind in units
of \msunyr.
\label{fig: disk3}
}
\end{figure}

Figure \ref{fig: disk3} shows the variation of disk mass and mass
accretion rate for the disk in Figure \ref{fig: disk2} and several
mass loss rates.  In general, the wind starts to empty the disk when
the mass accretion rate through the disk falls below the mass loss
rate in the wind \citep[see][and references therein]{alex2009, owen2010}. 
For very low mass loss rates, $\dot{M}_{wind} = 10^{-10}$ \msunyr, the 
wind evaporates the disk on timescales of 10 Myr
\citep[see also][2009]{alex2007}. As the mass loss rate from
photoevaporation grows, the disk evaporates on shorter and shorter
timescales. For $\dot{M}_{wind} = 10^{-8}$ \msunyr, the disk
disappears on timescales shorter than 1 Myr.

These evolutionary models capture the main observable features of
disks around pre-main sequence stars with ages of 1--10 Myr.  For 1
Myr-old young stars, disk masses of $\sim$ 0.001--0.1 \msun\ and mass
accretion rates of $1-100 \times 10^{-10}$ \msunyr\ are common
\citep[][2007]{gull1998,hart1998,and05}. Typical disk lifetimes are
1--3 Myr \citep{hart1998,hai01}. Few pre-main sequence stars have ages
larger than $\sim$ 10 Myr \citep{currie2007,mamajek2009,kenn09}. Thus, 
these calculations yield reasonable physical conditions for planet
formation.

In our previous calculations, we set the surface density of the gaseous 
disk as $\Sigma = \Sigma_0 x_m a^{-n} e^{-t/t_g}$, with $x_m$ as a scaling
factor, $n$ as a constant power law, and $t_g$ as a constant gas depletion 
time. In our new approach, adopting an analytic \citep{cha2009} or a 
numerical (eq.~\ref{eq: disk-evol}) solution to the radial diffusion equation 
has several important advantages for little computational effort.

\begin{itemize}

\item Following the evolution of the mass accretion rate onto the central star
enables robust comparisons between observations of accretion in pre-main sequence
stars with the timing of planet formation throughout the disk.

\item Treating the evolution of photoevaporation allows us to make links between 
the so-called transition disks and the formation of planets \citep[e.g.,][]{alex2009}.

\item Tracking the time evolution of the radial position of the snow line 
\citep[e.g.,][]{kenn08} in a physically self-consistent fashion gives us a way to 
include changes in the composition of planetesimals with time.

\item Calculating the radial expansion of the disk with time provides a way to 
compare the observed properties of protostellar disks \citep[e.g.,][2007, 2009]{and05}
with the derived properties of planet-forming disks.

\end{itemize}

It is straightforward to derive the properties of power-law disks that yield results 
similar to those of our new calculations. In analytic and numerical solutions to the
diffusion equation, the viscous time $t_{\nu} \approx R^2 / 3 \nu$ is roughly the
gas depletion time $t_g$. For the \citet{cha2009} analytic approach, 
$t_{\nu} \approx$ 4~Myr $(\alpha / 10^{-3})$; for our numerical solution, 
$t_{\nu} \approx$ 1~Myr $(\alpha / 10^{-3})$. In previous calculations, we adopt
$t_g$ = 10 Myr; thus, new calculations with $\alpha = 1 - 4 \times 10^{-4}$ 
roughly correspond to the disks in \citet[][2009, 2010]{kb08}. 

The new calculations generally yield shallower power law slopes -- $n$ = 0.6 for
\citet{cha2009} and $n$ = 1 for our numerical solution -- than the $n$ = 1.0--1.5
assumed in our previous studies. Because the timescale for planet formation is 
$t \propto a^{n + 1.5}$ \citep[see][and references therein]{kb08}, planets form
relatively faster in the outer disk in these new calculations. However, the range
of disk masses considered here, $M_{d,0}$ = 0.01--0.1 \msun, overlaps the
$M_{d,0}$ = 0.0003--0.25 \msun\ adopted in \citet{kb10}. Thus, calculations with 
similar initial disk masses and similar initial size distributions of planetesimals 
will yield similar timescales for the formation of the first oligarchs.

As a concrete example, we compare several disk surface density distributions 
quantitatively.  In \citet[][2009, 2010]{kb08}, we often adopted an initial
surface density distribution of solid material, 
$\Sigma$ = 30 g~cm$^{-2}$ $x_m$ (a / 1 AU)$^{-3/2}$.  At 5 AU, our numerical 
solutions for disks with $M_{d,0}$ = 0.1~\msun\ and $R_{d,0}$ = 30~AU have 
the same surface density as power law disks with $x_m$ = 3. For the same
initial disk mass and radius, the \citet{cha2009} analytic model has the same
surface density as power law disks with $x_m \approx$ 2.5.  For identical
initial conditions and equivalent viscous timescales, these three surface
density distributions yield similar evolutionary times for the growth of
protoplanets.

\subsection{Coagulation Code}

\citet{kb08} describe the main details of the coagulation code we use to 
calculate the growth of small solid particles (planetesimals) into planets. 
Briefly, we perform calculations on a cylindrical grid with inner radius
$R_{in}$ and outer radius $R_{out}$. The model grid contains $N$
concentric annuli with widths $\delta R_i$ centered at radii
$R_i$. Calculations begin with a mass distribution $n(m_{ik}$) of
planetesimals with horizontal and vertical velocities $h_{ik}(t)$ and
$v_{ik}(t)$ relative to a circular orbit.  The horizontal velocity is
related to the orbital eccentricity, $e_{ik}^2(t)$ = 1.6
$(h_{ik}(t)/V_{K,i})^2$, where $V_{K,i}$ is the circular orbital
velocity in annulus $i$.  The orbital inclination depends on the
vertical velocity, $i_{ik}^2(t)$ = sin$^{-1}(2(v_{ik}(t)/V_{K,i})^2)$.

The mass and velocity distributions of planetesimals evolve in time
due to inelastic collisions, drag forces, and gravitational forces. 
The collision rate is $n \sigma v f_g$, where $n$ is the number 
density of objects, $\sigma$ is the geometric cross-section, $v$ is 
the relative velocity, and $f_g$ is the gravitational focusing factor
\citep{saf69,lis87,spa91,wet93,wei97,kl98,kri06,the07,loh08,kb08}.
The collision outcome depends on the ratio of the collision energy
needed to eject half the mass of a pair of colliding planetesimals
\qdstar\ to the center of mass collision energy $Q_c$. If $m_1$ and
$m_2$ are the masses of two colliding planetesimals, the mass of the
merged planetesimal is
\begin{equation}
m = m_1 + m_2 - m_{ej} ~ ,
\label{eq:msum}
\end{equation}
where the mass of debris ejected in a collision is
\begin{equation}
m_{ej} = 0.5 ~ (m_1 + m_2) \left ( \frac{Q_c}{Q_d^*} \right)^{9/8} ~ .
\label{eq:mej}
\end{equation}
This approach allows us to derive ejected masses for catastrophic
collisions with $Q_c \sim Q_d^*$ and for cratering collisions with
$Q_c \ll Q_d^*$ \citep[see
  also][]{wet93,wil94,tan96,st97,kl99,obr03,kob10}. Consistent with
numerical simulations of collision outcomes
\citep[e.g.,][]{ben99,lein08,lei09}, we set
\begin{equation}
Q_d^* = Q_b r^{\beta_b} + Q_g \rho_g r^{\beta_g}
\label{eq:Qd}
\end{equation}
where $Q_b r^{\beta_b}$ is the bulk component of the binding energy,
$Q_g \rho_g r^{\beta_g}$ is the gravity component of the binding energy,
$r$ is the radius of a planetesimal, and $\rho_g$ is the mass density of
a planetesimal.

To compute the evolution of the velocity distribution, we include
collisional damping from inelastic collisions and gravitational 
interactions.  For inelastic and elastic collisions, we adopt the 
statistical, Fokker-Planck approaches of \citet{oht92} and \citet{oht02}, 
which treat pairwise interactions (e.g., dynamical friction and viscous 
stirring) between all objects in all annuli \citep[see also][]{ste00}.  
As in \citet{kb01}, we add terms to treat the probability that objects 
in annulus $i$ interact with objects in annulus $j$ 
\citep[see also][2008]{kb04b}. 

\subsubsection{Small particles}

In most of our previous calculations, we calculate the evolution of particles 
with radii larger than the `stopping radius,' $r_s \approx$ 0.5--2~m at 5--10 AU 
\citep{ada76,wei77,raf04}.  Although they are subject to gas drag, 
these particles are not well-coupled to the gas. Here, we also consider the 
evolution of smaller particles entrained by the gas \citep[see also][]{kb09}.  
Following \citet{you2007}, we derive the Stokes number
\begin{equation}
St = \frac{r \rho_g \Omega}{\rho c_s} ~ ,
\end{equation}
where $\rho = \Sigma / 2 H $ is the gas density. The vertical scale
height of small particles is then
\begin{equation}
H_s = \left\{ \begin{array}{cll}
      H ~ , &    & St \le \alpha \\
      \sqrt{\frac{\alpha}{St}} ~ H ~,  &    & St > \alpha \\
      \end{array} \right.
\label{eq:Hsmall}
\end{equation}
We assume that entrained small particles have vertical velocity $v = H_s \Omega$ 
and horizontal velocity $h = 1.6 v$.

Estimating accretion rates of small particles by embedded protoplanets is a challenging 
problem. For particles with $St \sim 1$, accretion rates depend on the strength of 
the planet's gravity relative to forces that couple particles to the gas. Recently, 
\citet{orm2010} began to address this issue with an analysis of interactions between
protoplanets and particles (loosely) coupled to the gas. After identifying three
classes of encounters, they derive growth rates as a function of $m$, $St$, and the
local properties of the gas. For massive protoplanets with $r \gtrsim$ 1000~km, they 
derive short growth times, $\sim 10^3$ yr, for particles with $10^{-1} < St < 10$.

Here, we adopt a simple prescription for accretion of small particles with 
$St \lesssim$ 1. Small protoplanets with $r < 100$ km do not have strong enough
gravitational fields to wrest small particles from the gas. Thus, we assume small
protoplanets cannot accrete particles with $St < 1$. The \citet{orm2010} results 
suggest large protoplanets accrete particles with $St < 10^{-3}$ on long timescales, 
$\gtrsim$ 1 Myr.  Thus, we also assume protoplanets of any size cannot accrete these 
particles. For particles with $St > 10^{-3}$, we assume accretion proceeds as in our 
standard formalism with collision rates $n \sigma v f_g$, where the cross-section 
$\sigma$ includes the enhanced radius of the protoplanet discussed in \S2.2.2. Our
approach generally yields growth rates a factor of 1--10 smaller than those of 
\citet{orm2010}. Thus, our formation times are longer than timescales derived from
a more rigorous approach.  For the calculations in this paper, however, this accuracy 
is sufficient to explore the impact of disk mass and viscosity on the formation of gas
giant planets.  In future papers, a more comprehensive treatment of small particle
accretion and gas drag will yield a better understanding of the role of the local
properties of the gas on planet formation.

Aside from enabling protoplanets to accrete smaller particles at large rates, 
including small particles with $r \lesssim$ 1~m allows us to derive a more accurate 
calculation for the time evolution of debris from the collisional cascade. Compared to 
results in \citet[][2010]{kb08}, estimates for the abundance of very small grains with
$r \sim$ 1--10 $\mu$m now rely on extrapolation over 3 orders of magnitude in radius 
instead of 6. Thus, this addition yields more accurate predictions for the variation
of grain emission as a function of time \citep[see, for example,][2010]{kb08}.

\subsubsection{Planetary atmospheres}

As planets grow, they acquire gaseous atmospheres. Initially, the
radius of the atmosphere $R_a$ is well-approximated by the smaller of
the Bondi radius $R_B$ and the Hill radius $R_H$:
\begin{equation}\label{eq:bondihillradii}
R_a = \left\{ \begin{array}{cll}
      R_B = \frac{G M_p}{c_s^2} ~ , &    & R_B < R_H \\
      R_H = \left ( \frac{M_p}{3 M_\star} \right )^{1/3} a ~ ,  &    & R_B > R_H \\
      \end{array} \right.
\label{eq: r-atmos}
\end{equation}
where $M_p$ is the mass of the planet and $a$ is the semimajor axis of
the planet's orbit around the central star. For low mass planets, $R_H
> R_B$. Requiring $R_B > R_P$, where $R_P$ is the radius of the
planet, an icy planet with a mass density of 1 \gcm3\ starts to form
an atmosphere when $M_p \gtrsim 3 \times 10^{25}$ g.

Once planets develop a small atmosphere, they accrete small particles
more rapidly \citep{pod88,kar93,ina2003,tan10}. As small particles 
approach the planet, atmospheric drag reduces their velocities. Smaller 
velocities increase gravitational focusing factors, enabling very rapid 
accretion \citep[see also][]{cha2008,raf2010}. Because the extent of the
atmosphere depends on the accretion luminosity, calculations need to
find the right balance between the extent of the atmosphere and the
accretion rate (and luminosity). Here, we follow \citet{ina2003}
and solve for the structure of an atmosphere in hydrostatic
equilibrium around each planet with $M_p \gtrsim 3 \times 10^{25}$~g. 
For each smaller planetesimal with mass $m$, we calculate the enhanced 
radius $R_E(M_p,~m)$, which replaces the physical radius in our derivation 
of cross-sections and collision rates.

This approach shortens growth times by factors of 3--10. In \citet{cha2006},
protoplanets with atmospheres grow 3--4 times faster at 5~AU than protoplanets 
without atmospheres \citep[see also][]{ina2003}.  To confirm these results, we
repeated the calculations of \citet{kb09} for protoplanets with atmospheres.
In a power-law disk with $x_m$ = 5, our previous calculations yielded 
10~\mearth\ protoplanets on timescales of $\sim$ 3 Myr. Repeating these calculations 
for protoplanets with planetary atmospheres results in growth times shorter 
by factors of 5--10, $\sim$ 0.3--0.5 Myr. Several numerical tests suggest
that different fragmentation laws produce factor of 1.5--2 changes in the
growth time for protoplanets with atmospheres \citep[see also][2008]{cha2006}. 
\citet{kb09} derived only 10\% to 20\% variations in growth times for a 
similar range in fragmentation laws.  Thus, growth times are more sensitive
to fragmentation when protoplanets have atmospheres.

\subsubsection{Gas accretion}

As planets continue to grow, they begin to accrete gas from the
disk. Although the minimum mass required to accrete gas varies with
the accretion rate of the planet, the properties of the disk near the
planet, and the distance of the planet from the central star, a
typical minimum mass is $\sim$ 1--10
\mearth\ \citep[][2010]{miz1980,ste1982,iko2000,hor2010,raf2006}.
Once gas accretion begins, the planet mass grows rapidly until it
removes most of the gas in its vicinity, either by depleting the disk
entirely or by opening up a gap between the planet and the rest of the
disk. The planet then grows more slowly and may start to migrate
through the disk \citep[e.g.,][]{lin1986,lin1996,ward1997,dang2010}.

Here, we approximate the complicated physics of gas accretion and the
evolution of the atmosphere with a simple formula 
\citep[see also][]{ver04,ida05,alex2009}. 
The growth rates reviewed in \citet{dang2010} suggest
\begin{equation}
\dot{M}_g \approx \dot{M}_0 \frac{M_p}{M_0} ~ \frac{\Sigma \Omega^2}{c_s} ~ e^{-((\mu - \mu_0)/\sigma_m)^2} ~ ,
\label{eq: mdotgas}
\end{equation}
where $\dot{M}_0$ is a typical maximum accretion rate, $M_0 \approx $
0.1 \mjup\ (1 \mjup\ is the mass of Jupiter), $\mu$ = log $M_p$,
$\mu_0$ = log $M_0$, and $\sigma_m \approx$ 2/3. This functional form
captures the rapid increase in the accretion rate from the minimum
core mass of 1--10 \mearth\ to larger core masses, the peak accretion
rate at masses of roughly 10\% the mass in Jupiter, and the decline in
the accretion rate once the planet opens up a gap in the disk.  Detailed 
numerical simulations suggest maximum accretion rates of 1 Jupiter mass 
every 0.01--0.1 Myr \citep[e.g.][and references therein]{lis07,dang2010,mach2010}.  
For simplicity, we consider $\dot{M}_0$ a free parameter; we set a rate
appropriate for a Jupiter mass gas giant in a disk with $\Sigma$ = 100~g~cm$^{-2}$, 
$\Omega = 1.78 \times 10^{-8}$ s$^{-1}$ (5 AU), and $c_s$ = 0.67 \kms\ and 
use equation (\ref{eq: mdotgas}) to scale this rate throughout the disk.

To remove accreted gas from the disk in our numerical simulations of
disk evolution, we set the change in surface density as
\begin{equation}
(\frac{\partial \Sigma}{\partial t})_{gas} = -\frac{\dot{M}_g}{2 \pi R \Delta R} ~ ,
\label{eq: dsigmagas}
\end{equation}
where $R$ is the semimajor axis of the planet and $\Delta R$ is the
width of the annulus.  Combined with photoevaporation from the central
star, this expression yields
\begin{equation}
(\frac{\partial \Sigma}{\partial t})_{ext} = \left\{ \begin{array}{cll}
      (\frac{\partial \Sigma}{\partial t})_{diffuse} + (\frac{\partial \Sigma}{\partial t})_{gas} ~ , &   & R_{in} \approx R_\star \\
      (\frac{\partial \Sigma}{\partial t})_{direct} + (\frac{\partial \Sigma}{\partial t})_{gas} ~ , &   & R_{in} \gg R_\star \\
      \end{array} \right.
\label{eq: dsigmatot}
\end{equation}
where $R_{in}$ is the inner edge of the optically thick portion of the disk.

Although we use equation (\ref{eq: mdotgas}) to derive gas accretion rates for the
analytic disk model, we do not remove the accreted mass from the disk. To place some
limit on the amount of mass accreted by gas giants, we halt gas accretion when the total 
mass in gas giants exceeds the total remaining mass in the disk. 

\subsubsection{Evolution of R and L of planets}

Throughout this phase, the radius of the planet $R_p$ depends on the accretion rate and the 
properties of the atmosphere \citep[e.g.,][]{hart1997,pap2005,cha07,lis07,bar2009,dang2010}. 
Before the planet opens a gap in the disk, accretion is roughly spherical; after the 
gap forms, the planet accretes material from a thin disk-like structure within the 
planet's Hill sphere \citep[e.g.,][]{nel2000,dang2008}.
For planets with masses much larger than Neptune, disk accretion provides most of the 
gas. Thus, to derive a simple estimate for the evolution of $R_p$, we solve the energy 
equation for a polytrope of index $n = 1.5$ accreting material from a gaseous disk 
\citep[e.g.][]{hart1997}:
\begin{equation}
\frac{\dot{R_p}}{R_p} = \left ( \frac{7}{3} ~ \eta - \frac{1}{3} \right ) \frac{\dot{M_p}}{M_p} ~ - ~ \frac{7}{3} \left ( \frac{L}{G M_p^2/R_p} \right ) ~ .
\label{eq:rdot}
\end{equation}
Here, $L$ is the planet's photospheric luminosity in erg s$^{-1}$; 
$1 - \eta$, where $\eta \le 1$, measures the fraction of the accretion energy 
radiated away before material hits the planet's photosphere\footnote{To avoid 
confusion with our use of $\alpha$ for the disk viscosity parameter, we change 
the $\alpha$ in \citet{hart1997} to $\eta$.}. 

Solving equation (\ref{eq:rdot}) requires a relation for $L$.  When $\dot{M}$ = 0, planets 
resemble $n$ = 0.5--1 polytropes, with $R_p \propto M_p^{m}$ and $-1/8 \le m \le 1/10$ 
\citep[e.g.,][]{sau1996,fort2009}.  For this phase, we derive a simple expression from 
published evolutionary tracks \citep[e.g.,][2008]{burr1997,spi2010,bar2003}:
\begin{equation}
L \approx 10^{-9} ~ \lsun ~ \left ( \frac{M_p}{1~M_J} \right )^{0.25} ~ \left ( \frac{R_p}{1~R_J} \right )^{15} , ~
\label{eq:lp1}
\end{equation}
where $R_J$ is the radius of Jupiter.  Accretion energy tends to expand planets considerably 
\citep[e.g.,][]{pol96,mar2007,dang2010}. To treat this phase, we derive a simple expression 
from published model atmosphere calculations applied to an $n = 1.5$ polytrope 
\citep[][and references therein]{sau1996,fort2009}:
\begin{equation}
L \approx 10^{-6} ~ \lsun ~ \left ( \frac{M_p}{1~M_J} \right ) ~ \left ( \frac{R_p}{1~R_J} \right )^{3} . ~
\label{eq:lp2}
\end{equation}
When the planet has a radius $R_t \approx 1.78 (M_p/1~M_J)^{0.06} R_J$, these relations yield the 
same luminosity. Thus, we use equation (\ref{eq:lp1}) for $R(M_p) > R_t$ and $\dot{M}_g >$ 0 and 
equation (\ref{eq:lp2}) otherwise.

\begin{figure}
\includegraphics[width=6.5in]{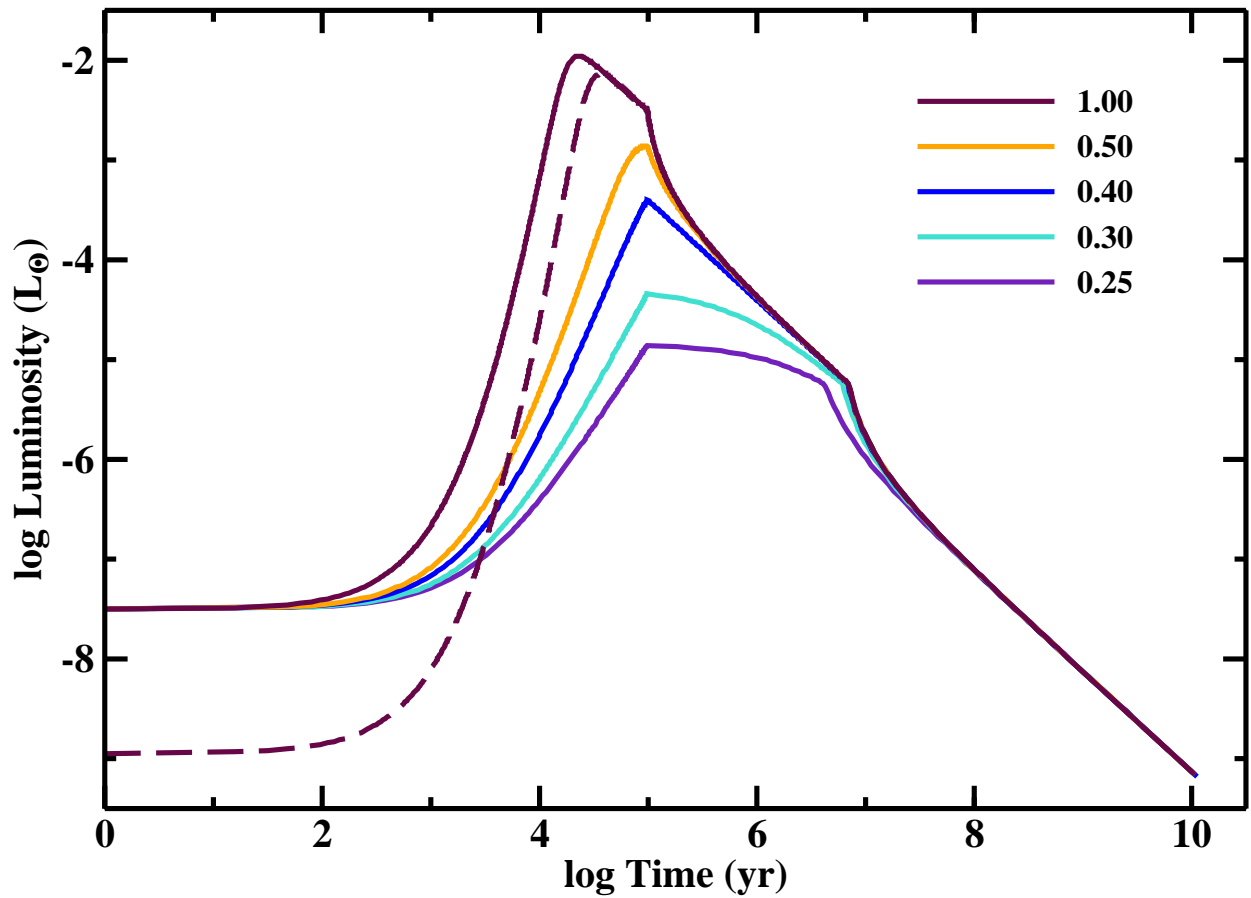}
\vskip 3ex
\caption{
Luminosity evolution for a gas giant planet using the prescription in equations
(\ref{eq:rdot})--(\ref{eq:lp2}). Starting with an initial mass of 10 \mearth\ and 
an initial radius of 1 \rjup, the planet accretes gas at a rate of $10^{-5}$ 
\mjup\ yr$^{-1}$ until it reaches a mass of 1 \mjup.  Solid curves show the evolution 
for various $\eta$.  When $\eta$ is small; the planet accretes cold material and
has a small peak luminosity. When $\eta$ is large, the planet accretes hotter
material, expands, and has a large peak luminosity. At late times, all curves
converge on a single evolutionary track which yields a Jupiter-radius planet 
at 4--6 Gyr. The dashed curve illustrates how the evolution changes for planets 
with initial radii of 0.33 \rjup\ (comparable to the radius of Neptune).
}
\label{fig: levol}
\end{figure}

\begin{figure}
\includegraphics[width=6.5in]{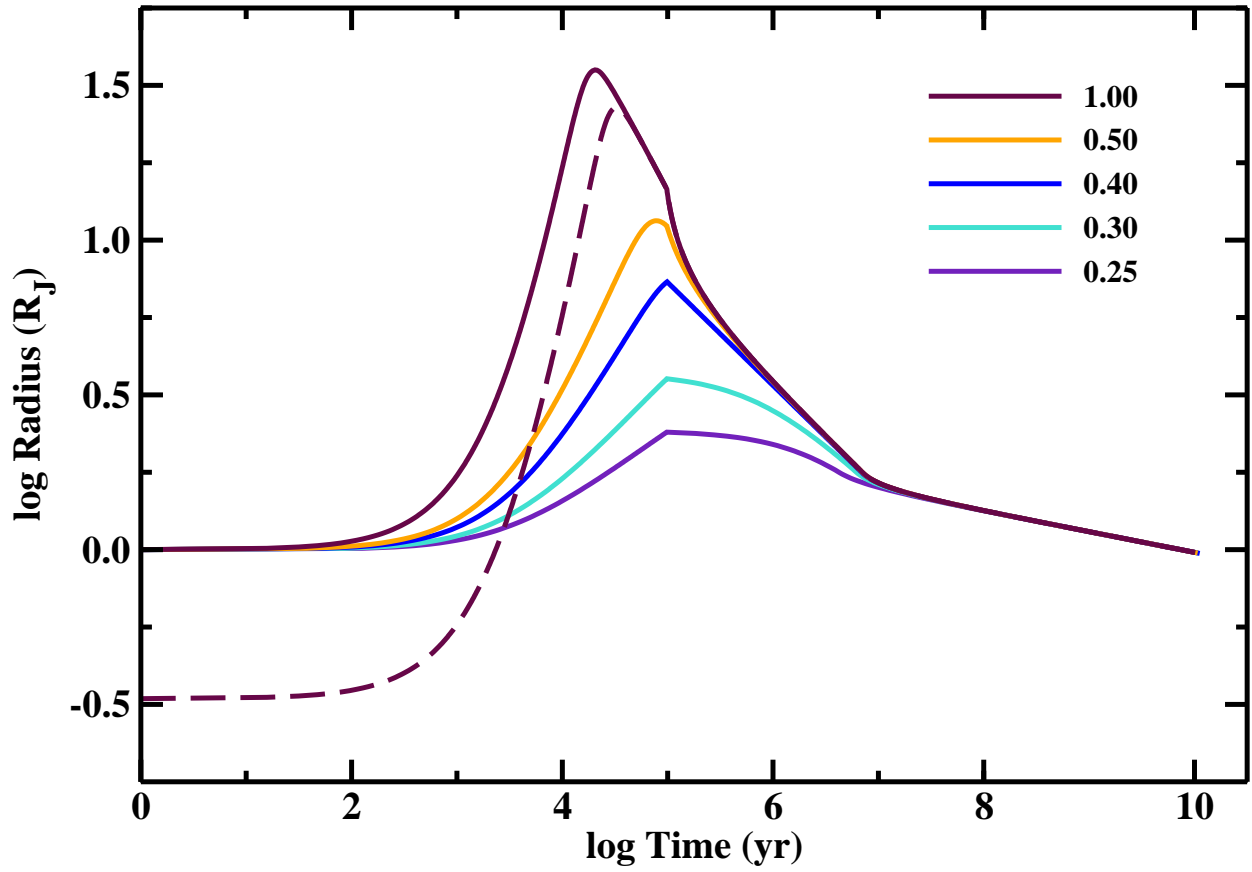}
\vskip 3ex
\caption{
As in Figure \ref{fig: levol} for the radius of the planet. Planets that accrete
hotter gas grow to larger radii.
}
\label{fig: revol}
\end{figure}

Despite its simplicity, this approach yields reasonable results for $L(t)$ and $R_p(t)$. In the 
example of Fig. \ref{fig: levol}--\ref{fig: revol}, the calculation starts with a 10 \mearth, 
1 \rjup\ planet accreting at a rate of $10^{-5}$ \mjupyr\ until it reaches a mass of 1 \mjup.  Before 
the planet reaches its final mass, the luminosity increases exponentially with time.  Once accretion 
stops, the luminosity declines. Peak luminosity and radius depend on $\eta$; planets that accrete 
hotter material (larger $\eta$) expand more than planets that accrete colder material (smaller $\eta$).
For $\eta \approx$ 0.3--0.4, our peak luminosities of $\sim 10^{-4}$ \lsun\ are similar to those
of more detailed calculations \citep{pol96,mar2007,dang2010}. At late times, evolution for all $\eta$ 
converges on a single track for $L(t)$ and $R_p(t)$. This track yields a radius of 1.03 \rjup\ at 4.5 Gyr.

Although this approach is not precise, it serves our purposes.  The model yields planetary radii 
with sufficient accuracy ($\pm$ 10\%) to derive the merger rate of growing planets from dynamical
calculations with our $N$-body code. The estimated luminosity is also accurate enough to serve as 
a reasonable starting point for more detailed evolutionary calculations, as in \citet{burr1997} or 
\citet{bar2003}. This formalism is also flexible: it is simple to adopt better prescriptions for
the luminosity or an energy equation with a different polytropic index.

\subsection{$N$-body Code}

\citet{bk06} provide a description of the $N$-body component of our hybrid code.  
To solve the equations of motion for a set of interacting particles, we use an 
adaptive algorithm with sixth-order time steps, based either on Richardson 
extrapolation \citep{bk06} or a symplectic method \citep{yos90,kin91,wis91,sah92}.
The code calculates gravitational forces by direct
summation and evolves particles accordingly.  In our earlier version,
we performed integrations in terms of phase-space variables defined
relative to individual Keplerian frames, following the work of Encke
\citep{enc52,vas82,wis91,ida92,fuk96,she02}; our latest version can
track center-of-mass frame coordinates instead. This feature may be
useful in instances where stellar recoil is important.  As before, the
code evolves close encounters -- including mergers -- between pairs by 
solution of Kepler's equations in the pairs' center of mass frame. The
algorithm also includes changes in eccentricity, inclination, and 
semimajor axis, derived from the Fokker-Planck and gas drag algorithms
in the coagulation code.

The 2006 version and the new version of $N$-body code can track the force
from a dusty or gaseous disk. The older code handles only toy models for 
the gaseous disk, consisting of a power law surface density profile and 
an exponential decay in time.  The new code includes the more realistic disk
potential derived in \S\ref{subsect:diskevolve}. The code represents a disk 
in annular bins with time-varying surface density specified by the physical 
model. Although the current code does not treat the lack of gravitational 
force from gaps in the disk produced by gas giant planets, this extra force
is small compared to the forces from the rest of the disk and gas giant planets.
Once this feature is included, we can then self-consistently track the 
dynamics of gas giants in an evolving gaseous disk.

Appendix~\ref{appx:disk} provides some background and details of our method
to treat the gravity of the gaseous disk. To summarize, we approximate the 
disk as axisymmetric with a scale height $H$, which is small compared to the 
orbital distance ($H/a \simless 0.1$). The acceleration that the disk produces 
on a planet near the disk midplane is then fast to calculate. When the disk 
has a power-law surface density, a simple analytical expression suffices.  If 
the disk has more complicated radial structure (\S\ref{subsect:diskevolve}),
then we split it up into constant-density annuli and calculate the contribution 
of each annulus to the acceleration at some point near the disk midplane. We 
repeat this calculation for a set of $O(10^4)$ radial points at the beginning 
of a simulation, and interpolate with a cubic spline thereafter to find the 
disk acceleration at the location of the $N$ objects.  As the surface density 
of the disk evolves, the code updates the date for the spline fit.

In addition to gas, the new version of the $N$-body code includes the gravity
of non-physical objects representing dust or planetesimals.  As in the SyMBA 
code of \citet{dun98}, we use massless ``tracer'' particles, which simply respond 
to the gravitational potential produced by the central star and planets.  We use 
the evolution of the tracers to inform the coagulation code of the $de/dt$, 
$di/dt$, and $da/dt$ induced by massive planets in the $N$-body code.  We also 
have a second population of massive ``swarm'' particles, which have mutual
interactions with planets but do not interact with either tracers or with each 
other. The swarm thus consists of super-particles which represent the ensemble
of small objects that interact with each other statistically in the coagulation 
code and which also interact dynamically with planets in the $N$-body code.
Independently of the tracers, we can use the swarm particles to inform the 
coagulation code of the $de/dt$, $di/dt$, and $da/dt$ induced by massive planets 
in the $N$-body code. More importantly, we use the mutual interactions of swarm
and $N$-body particles to calculate the response of planets to the surface density
and velocity distributions of the swarm. In addition to the tests in \S3, 
\cite{bk11} describe some applications of the interaction between swarm and
$N$-body particles.

In terms of comparative computational cost, planets are expensive, tracers are cheap, 
and swarm particles are in between.  Even without mutual interactions, the swarm 
particles can be a considerable burden in long-term integrations. To lighten this 
computational load, we evolve the star and planets independently of the swarm during 
a single coarse-grain timestep, allowing the code to take as many substeps as necessary.  
Then, we interpolate the resulting planet trajectories to calculate the gravitational 
forces needed to update the swarm. When we use massive swarm particles, we allow the 
planets to deviate from their interpolated path in response to the smaller objects.  
This approach is valid only when the net forces on the planets from the swarm vary 
slowly in time, and when the our simple third-order interpolation of the planet's 
trajectory over a coarse timestep is realistic. With this approach, there is a risk 
of lower accuracy for the very rare events when some members of the swarm interact 
with a pair of closely-interacting $N$-body particles. Our 3rd-order interpolation 
then does not yield a robust representation of the trajectories of these swarm 
particles. Detailed tests show that reduced accuracy only occurs for the very few 
swarm particles within 1--2 Hill radii of the pair of $N$-bodies, has a negligible 
impact on the trajectory of the $N$-body particles, and does not influence the 
outcome (merger or scattering event) of the interaction between the two $N$-bodies.

\subsection{$N$-body code tests}\label{subsect:nbodytests}

We put the new version of the $N$-body code through the same
diagnostics as in \citet{bk06}. To assess accuracy and dynamic range,
we perform long-term orbit integrations of the major planets, as well
as integrations of the planets with their masses scaled by a factor of
fifty \citep{dun98,bk06}. We also track the motion of tight binaries
in orbit around a central star. In one limit, we evolve a close pair of
Jupiter-mass objects \citep{dun98}; in another, we simulate a surface-skimming 
satellite above the Earth as it orbits the Sun. We also set up pairs of 
planets on closely-spaced circular orbits and evolve them to see whether we can 
resolve the critical separation that determines whether their orbits
cross\footnote{Our code yields a value within a few percent of the
predicted critical separation, $\Delta a_{crit} = 2\sqrt{3} R_{H}$,
where $R_H$ is the mutual Hill radius (as in
eq.~\ref{eq:bondihillradii}, but with $M_p$ set to the sum of the
planets' masses).} \citep{gla93}.
With these tests we verify that the new code can replicate the results
illustrated in Figures~1--4 of \citet{bk06}.

\begin{figure}
\includegraphics[width=6.0in]{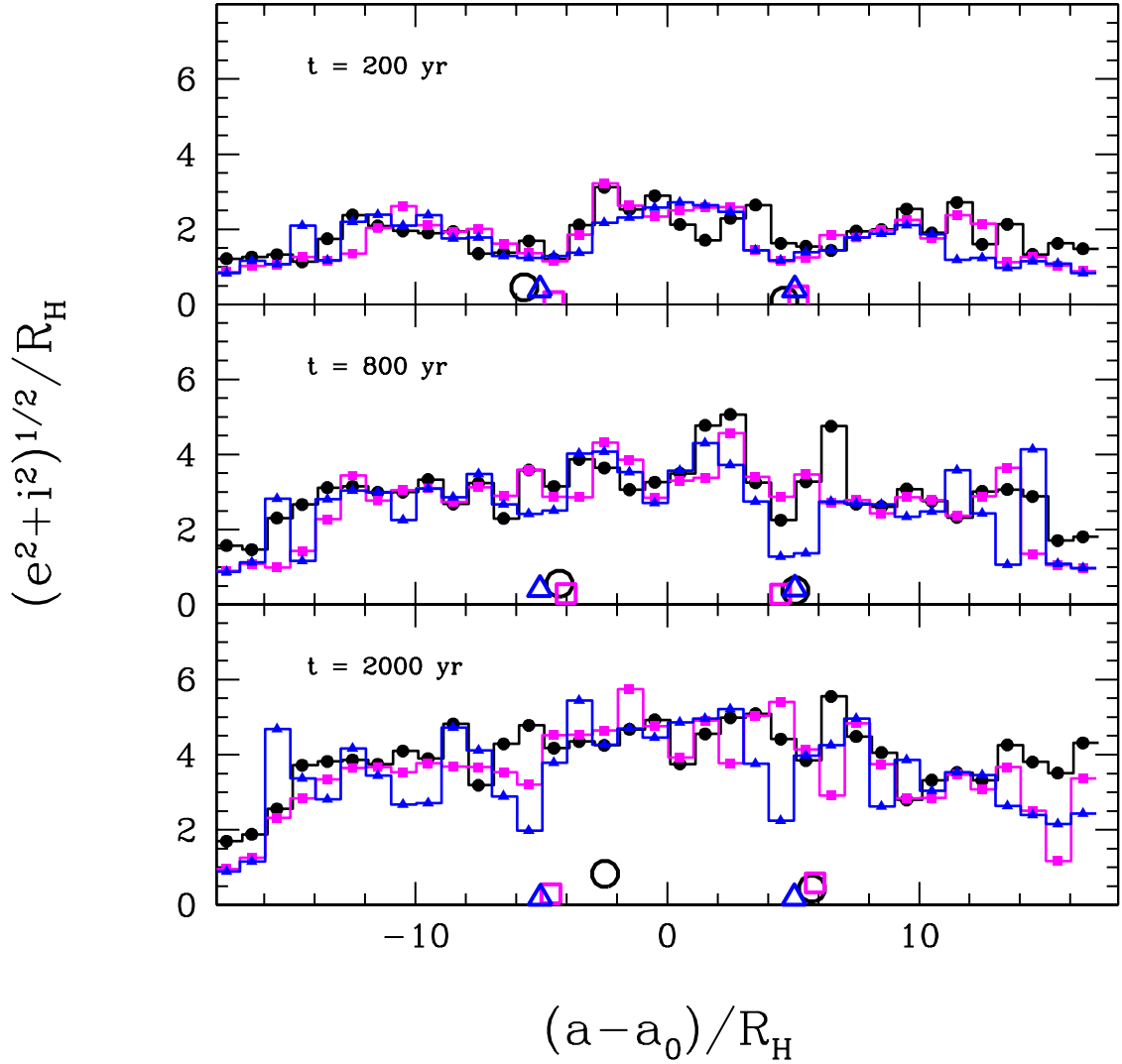}
\vskip -3ex
\caption{The evolution of $(e^2+i^2)^{1/2}$ (in units of Hill radii)
  of a particle disk with two embedded planets, as in
  \citet{kok95}. The abscissa gives the relative orbital separation 
  in terms of the difference between the semimajor axis $a$ and a
  reference radius $a_0 = 1$~AU. The histograms correspond to the
  disk particles; the symbols represent the two planets. The colors 
  and symbol styles distinguish simulations: black histograms
  and circles indicate fully interacting massive particles, blue lines
  and triangles show data from a massive disk of swarm particles with no
  self-gravity, while magenta histograms and squares correspond to a
  massless disk composed of tracers.
\label{fig:stir}
}
\end{figure}

To save computation time, the new code evolves tracers and swarm particles with 
lower temporal resolution than with the planets.  To verify that the resulting 
orbits are reasonable, we follow \citet{kok95} in simulating gravitational
stirring of a planetesimal swarm by a pair of planets. Two
$2\times 10^{26}$~g planets are separated by 10~$R_H$ and are embedded
in the middle of an annulus of 800 $2\times 10^{24}$~g objects, which
is 35~$R_H$ wide and centered at a distance of 1~AU from a 1~\msun\ star. 
All objects are initially on circular orbits. We evolve this
system treating all particles as mutually interacting massive
bodies, then with the disk composed of massive swarm particles
that interact with the planets only, and finally with a massless
tracer disk.  Figure~\ref{fig:stir} shows that all three methods yield
output that is in good qualitative agreement with previous
calculations \citep{kok95,wei97,kb01,bk06}.

The final set of tests for the $N$-body code concerns its ability to
identify mergers, in particular between planets and swarm particles or
tracers.  We first confirm that minor changes to the code maintain its
accuracy in high-speed mergers between small massive objects
\citep[the ``grapefruit test'' in][]{bk06}.  We then test mergers
between planets and tracers or swarm particles explicitly with the
more sophisticated method proposed by \citet{gre90}.  We set up (i) a 
$10^{-6}$~\msun\ planet on a circular orbit at 1~AU and (ii) test particles 
on orbits with eccentricity $e = 0.007$, inclination $i=0.2$~degrees, 
and semimajor axes distributed in two rings (0.977~AU to 0.991~AU and 
1.009~AU to 1.023~AU). We evolve the system through a single close 
encounter between the planet and each test particle to measure the 
fraction of test particles accreted by the planet. With the planet's 
physical radius set to 5300~km, we derive a merger fraction of 
$0.008\pm 0.001$, consistent with previous estimates \citep{gre90,dun98,bk06}.

\section{MIGRATION}

Interactions between a gaseous or planetesimal disk and planets can
lead to signification migration, with rates of change in semimajor
axis, $da/dt$, that are important on planet formation timescales
\citep{lin79,gol80,lin1986,ward1997,ver04,lev07}. Migration occurs when a planet 
perturbs a disk gravitationally. The resulting angular momentum exchange 
between the planet and the perturbed disk causes the planet to move radially 
inward or outward.  Several types of migration \citep{ward1997} depend on 
disk and planet properties.  When an embedded planet produces a modest 
perturbation in a massive disk (type I migration), the planet responds to 
torques from the disk and migrates inwards. When a planet is massive enough 
to open a radial gap in the disk (type II migration), the planet becomes 
locked into the global viscous flow and migrates inwards.  Type III migration 
\citep[e.g.,][]{pap07} is a fast migration mode associated with smaller 
planets that efficiently exchange angular momentum with co-orbiting material.
As with migration in a planetesimal disk (see Fig.~\ref{fig:nep}), this mode
can produce inward or outward migration.

Our code treats all types of migration through the gaseous disk using 
semianalytical approximations \citep[e.g.,][]{pap07}. Although we do not 
include any detailed disk hydrodynamics, the code can smoothly change the 
semimajor axis and other orbital elements of each planet at any specfied rate. 

The code includes another migration mechanism, perhaps ``Type 0,'' that 
involves interaction between a massive disk and planets. In an axisymmetric
disk that redistributes mass, planets migrate as the disk mass changes.
For photoevaporation and viscous transport, the disk mass and gravitational
potential change slowly compared to the local dynamical time. The angular
momentum of the planet is conserved.  By treating the disk and the central 
star as a point mass with mass $M_{eff}$ acting on a planet at semimajor 
axis $a$, $a M_{eff}$ is conserved. As the disk vanishes, the planet migrates
outward. We provide a few more details of Type 0 migration in Appendix~\ref{appx:type0}.

Figure~\ref{fig:type0} illustrates Type 0 migration in a disk with a power law
surface density distribution ($\Sigma \propto a^{-n}$) that decays 
exponentially on a timescale of $10^4$~yr. Two of the disks have $n = 1$; 
two others have $n = 1.5$. Each pair has an inner edge at 0.1 AU and an outer 
edge at either 50~AU or 100~AU. In all cases, the initial disk mass is 20\% of 
the mass of the 1~\msun\ central star. The migration scales approximately
with the amount of disk mass inside a planet's orbit. Therefore the
effect is strongest for a steep slope in surface density and smaller outer 
disk radius. While this type of migration may not be comparatively strong 
in magnitude, it may help to alter the orbital dynamics in a disk with
closely-packed giant planets. 

Type 0 migration pushes planets outward if the disk photoevaporates. It may push 
planets inwards if mass is flowing in toward the star by viscous processes.

\subsection{Migration in a planetesimal disk}

As protoplanetary disks evolve, planetesimals become important to the large-scale 
orbital dynamics of the growing planets. Repeated weak interactions with 
planetesimals can push a planet radially inwards or outwards. \citet{lin79} and
\citet{gol80} first quantified aspects of this effect analytically; since then,  
others have explored it numerically \citep[e.g.,][]{mal93,ida00,kir09}. Here, we
show that our code can perform similar calculations.

Our first illustration of planet migration follows \citet{kir09}. A 2~\mearth\ planet 
is embedded in a disk of planetesimals, each having a mass $1/600^{\rm th}$ of the 
planet's mass.  The disk extends from 14.5~AU to 35.5~AU and has a surface density 
$\Sigma = 1.2 ~ (a/a_0)^{-1}$ g~cm$^{-2}$ where $a_0$ = 25 AU.  Initially, the planet has 
$e$ = 0; the planetesimals have $e_{rms} = 0.001$ and $i_{rms}$ = 0.5 $e_{rms}$, where 
'rms' is root-mean-squared.  The planet stirs nearby planetesimals to large $e$
and ejects some planetesimals from the system (Figure~\ref{fig:migrate_ae}).  As a 
reaction to the stirring by the planetesimals, the planet migrates from 25 AU at 
$10^4$~yr to 22~AU at $6 \times 10^4$~yr, leaving excited planetesimals in its wake.
Because the planetesimals have finite masses and encounter the planet sporadically,
inward migration is not smooth (Figure~\ref{fig:migrate_ap}). Our results agree with
previous simulations \citep[see Figs.\ 3 and 4 of][]{kir09}.

Our second demonstration of migration follows the work of \citet{hah99}. Their disk
has $\Sigma \propto a^{-1}$ and total mass of 100~\mearth\ between 10~AU and 50~AU.
The disk is composed of 1000 planetesimals with $e_{rms} = 2 i_{rms} = 0.01$. With 
the Solar System's four major planets initially packed between 5~AU and 23~AU, the 
system evolves dramatically in time.  Dynamical interactions between the planets and 
planetesimals spread the orbits significantly. Jupiter migrates inwards a small
amount; the outer planets migrate outwards in semimajor axis. After $\sim$ 30~Myr,
Neptune reaches $a \sim 40$~AU. Figure~\ref{fig:nep} shows the specifics of the 
migration with our code. The results compare well with \citet[][lower left panel 
of their Figure~4]{hah99}.

\section{FORMATION OF GAS GIANT PLANETS}

To illustrate how gas giants form in our approach, we adopt the analytic disk 
model of \citet{cha2009} with $R_{d,0}$ = 30 AU, $M_{d,0}$ = 0.01, 0.02, 0.04, 
or 0.1 \msun, and $\alpha$ = $10^{-2}$, $10^{-3}$, $10^{-4}$, or $10^{-5}$. 
We follow the evolution of disk properties ($\Sigma$, $\dot{M}(R)$, $T_d$, etc)
on a grid extending from 0.2 AU to $10^4$ AU. For the coagulation code, we 
divide the region between 3 AU and 30 AU into 96 concentric annuli, adopt a 
dust-to-gas ratio of 0.01, and seed these annuli with planetesimals. To set the
gas parameters in the coagulation grid, we interpolate physical variables in 
the grid for the disk evolution code. For the bulk properties of planetesimals, 
we follow \citet{lei09} and adopt parameters for weaker objects with $f_W$ = 
\{$Q_b$ = $2 \times 10^5$ erg g$^{-1}$ cm$^{0.4}$, $\beta_b = -0.4$, $Q_g$ = 
0.33 erg g$^{-2}$ cm$^{1.7}$, $\beta_g$ = 1.3\} \citep[see][and references therein]{kb10}. 

In all 
theories, the mechanism, timing, and size distribution of planetesimal formation 
is uncertain \citep[e.g.,][]{ric06,gar07,kre07,bra08,cuz08,lai08,bir09,chi10,you2010}.
To explore how the formation of large planetesimals by the streaming instability 
\citep[see][and references therein]{you2010}
might produce gas giants, we consider two extreme possibilities.

\begin{enumerate}

\item Each annulus contains a single large `seed' planetesimal, $r \approx$ 1000 km, 
and a large population of 0.1--10 cm pebbles. The gas entrains small planetesimals.
The growth rate of the seeds then depends on the scale height of the gas, which
is set by the viscosity parameter $\alpha$. Because disks with smaller $\alpha$
have smaller vertical scale heights, planets should form more efficiently in 
disks with smaller $\alpha$. 

\item Each annulus contains only large planetesimals, $r \approx 1000$ km,
with initial $e = 10^{-4}$.  Aside from migration, the gas 
does not affect large planetesimals. However, collision rates scale inversely 
with the radius of planetesimals \citep[e.g.,][]{gol04,kb08}. Thus, planets should
form less efficiently than in calculations with a few large seeds and a swarm of 
small pebbles.

\end{enumerate}

For each of these possibilities, the coagulation code treats the evolution of 
planetesimals with radii of 1~mm to 3600~km. At the start of each calculation,
we assign a seed to the random number generator used to assign integral collision
rates \citep[e.g.,][]{wet93,kl98}. In this way, otherwise identical starting conditions 
can lead to very different collision histories.  As long as the gas density is
sufficient to entrain particles with $r \gtrsim$ 1 mm, we distribute collision
fragments with $r < $ 1~mm into mass bins with $r$ = 1--2 mm. Otherwise, these
particles are lost to the grid. When individual coagulation particles reach masses
of $m_{pro} \approx 3 \times 10^{26}$ g ($r \approx$ 3600~km), the code promotes
these objects into the $N$-body code. After the coagulation code initializes $a$, 
$e$, $i$, and the longitude of periastron, the $N$-body code follows the changing 
masses and orbital parameters of promoted objects. Although the $N$-body code can
treat the evolution of $N$-bodies with any orbital semimajor axis, we assume that
objects with $a \gtrsim 10^4 - 10^5$ AU are ejected from the system.

In these calculations, we adopt a single set of parameters for gas accretion onto icy 
cores. Throughout the disk, gas accretion begins when the core mass is 10~\mearth\ and 
the radius of the planet is 0.3~\rjup.
Gas is accreted cold ($\eta$ = 0.15). Thus, planets are small; mergers of gas giants
are relatively uncommon.  Planets accrete gas fairly slowly, with $\dot{M}_0 = $ 
2.0 \mjup\ Myr$^{-1}$, $M_0 = $ 0.1~\mjup, and $\sigma_m = $ 2/3. Because planets 
remain small, they have typical luminosities 1--2 orders of magnitude smaller than  
gas giants in other simulations \citep[e.g.,][]{hub05,lis07,dang2010}.

To isolate the importance of disk physics on our results, we ignore gas drag, migration,
and photoevaporation. For the properties of the disks and planetesimals we consider, 
growth times are generally faster than drag times. In our models, growing protoplanets 
orbit in a sea of relatively high eccentricity planetesimals until they begin chaotic
growth. Thus, the conditions required for rapid migration through a sea of planetesimals 
are never realized \citep{kir09,bk11}. 

Many analyses demonstrate that migration through the gas and photoevaporation are important 
processes in arriving at the final distribution of masses and orbital parameters for gas 
giants \citep[e.g.,][and references therein]{ida05,ali05,pap07,tho08,alex2009,lev10}.  In 
Kenyon \& Bromley (2011a, in preparation), we show how photoevaporation sets the maximum 
masses for gas giants as a function of $\alpha$ and $M_0$.  By analogy with numerical 
calculations of planets in disks of planetesimals, \cite{bk11} speculate that growing 
protoplanets are packed too closely to undergo type I migration \citep{ward1997,tan02} 
through the gas until they begin chaotic growth. Here, we concentrate on understanding 
how the growth of planets depends on initial disk properties.  Kenyon \& Bromley (2011b, 
in preparation) will address how the various types of migration change predictions for 
the masses and orbits of gas giant planets.

To summarize, our algorithms follow the growth and evolution of planets on three separate 
grids centered roughly on the central star. We derive a solution for the evolution of the
gaseous disk on a 1-D grid extending from 0.2~AU to $10^4$~AU.  As the most time-consuming 
part of the calculation, we follow the evolution of small solid objects on a smaller,
axisymmetric 2-D grid extending from 3--30 AU. Once the coagulation code promotes objects,
the $N$-body code follows their trajectories on a 3-D grid extending from the central star
to $\sim 10^4 - 10^5$ AU. Although objects can accrete solid material only from annuli
at 3--30 AU, giant planets can accrete gas from 0.2~AU to $10^4$~AU.  To save computer
time for a large set of calculations with identical starting conditions, we halted each
calculation at an evolution time of $\sim$ 10 Myr. In future papers, we will investigate
the long-term stability of planetary systems with gas giant planets.

\subsection{Calculations with Ensembles of Plutos}

We begin with a discussion of planet formation in a disk composed of gas and Pluto-mass
planetesimals. In these calculations, disk evolution scales with $\alpha$. Disks with
$\alpha = 10^{-2}$ evolve rapidly. For $M_{d,0}$ = 0.1~\msun, the disk mass declines to
0.07~\msun\ in 0.1~Myr, 0.03~\msun\ in 1~Myr, and 0.008~\msun\ in 10~Myr. Disks with
$\alpha = 10^{-5}$ evolve slowly. For all initial masses, the disk mass declines by 
less than 1\% in 10 Myr. 

In disks with ensembles of Plutos, solid objects grow very slowly. In a disk with 
$M_{d,0}$ = 0.1 \msun, it takes only $\sim$ 50--100 yr for the first Pluto-Pluto 
collision at $\sim$ 3 AU. Due to mutual viscous stirring, this first collision occurs 
at a modest velocity, $e \approx 10^{-3}$. Thus, this collision yields an object with 
a mass of roughly twice the mass of Pluto and some smaller collision fragments. 
Despite this promising start, it takes $\sim$ 0.1 Myr for this object to accrete another 
10 Plutos and to reach a mass of roughly $10^{26}$ g. Although these large objects accrete 
the collision fragments efficiently, the fragments have a very small fraction, $\sim$ 
0.1--1\%, of the initial mass. Thus, accretion proceeds solely by the very slow collisions 
of large objects. 

The dynamical evolution of the growing Plutos is also very slow.  At $\sim$ 0.5 Myr, the 
first objects reach masses of $\sim 3 \times 10^{26}$ g and are promoted into the $N$-body
code. With typical orbital separations of $\gtrsim$ 0.1 AU, these objects are spaced at 
intervals of $\gtrsim$ 10 Hill radii. Thus, they do not interact strongly. After $\sim$ 
2--3 Myr, the largest objects have masses of 0.1--0.5 \mearth\ and typical orbital
separations of 0.1--0.2 AU. A few of these are close enough to interact gravitationally.
However, the interactions are slow. With masses much less than an Earth mass, these objects 
sometimes reach masses of 0.5--1 \mearth\ after 10--20 Myr. These objects will never 
accrete gas and become gas giant planets.

Figure \ref{fig:pluto-prob} shows the mass distributions for these calculations at 3 Myr.
Independent of $\alpha$, the maximum planet mass scales with the initial disk mass as
\begin{equation}
M_{max} \approx 0.5 ~ \mearth\ \left ( \frac{M_{d,0}}{0.1~M_\odot} \right )^{-1.5} ~ .
\label{eq: mmax1}
\end{equation}
Because the most massive planets result from several rare collisions, the median mass 
is much smaller, $\sim$ 0.05 \mearth\ for disks with $M_{d,0} \approx$ 0.02--0.1 \msun.  
Disks with smaller initial masses cannot produce planets more massive than $\sim$ 0.01 \mearth. 

For massive disks, the mass distribution is roughly a power law, with a cumulative
probability $p \propto m^{-\beta}$. Our results suggest that more massive disks 
have shallower power laws, with $\beta \approx$ 2 for $M_{d,0} =$ 0.1 \msun\ and
$\beta \approx $ 4 for $M_{d,0} =$ 0.04 \msun.

These calculations demonstrate that disks composed of Pluto-mass objects can never 
become gas giant planets. This result is not surprising. For a disk with surface
density $\Sigma$, the accretion time is roughly $t \propto R P / \Sigma$, where
$P$ is the orbital period \citep[e.g.][]{lis87,gol04}. Thus, large planetesimals 
grow more slowly than small planetesimals \citep[see also][]{kb08}. In the limit 
of large objects growing without gravitational focusing, equation (56) of \citet{gol04} 
suggests that an ensemble of Plutos can produce 10 \mearth\ objects in 200--400 Myr
at 3 AU. Our numerical simulations confirm this long growth time.

\subsection{Calculations with Pebbles and a few Plutos}

Disks composed of a few Plutos and many pebbles evolve rapidly. For a model with
$M_{d,0}$ = 0.1 \msun, $R_{d,0}$ = 30 AU, and $\alpha = 10^{-3}$, it takes $\sim$ 
1.5--1.8~Myr to promote the first object into the $N$-body code. Within 0.5 Myr,
another 10--15 objects reach the promotion mass of $3 \times 10^{26}$~g.  These 
objects have small atmospheres; they rapidly accrete the remaining pebbles in their 
feeding zones. It typically takes $\sim$ 0.2~Myr for protoplanets to reach masses 
of 1--2~\mearth\ and another 0.2--0.5~Myr to begin to accrete gas from the 
disk. Within a total evolution time of 2.5--3~Myr, some protoplanets grow into gas
giant planets.

The disk viscosity parameter $\alpha$ sets the timescale for the early portions of
this evolution. In calculations with $\alpha = 10^{-2}$, the timescale to produce
the first $N$-body is $\sim$ 2~Myr, much longer than the 0.2--0.4 Myr timescale for 
calculations with $\alpha = 10^{-4} - 10^{-5}$. In these models, $\alpha$ sets the
scale height of the pebbles (equation (\ref{eq:Hsmall})). Gravitational focusing 
factors grow with smaller scale heights; thus, the Pluto-mass seeds accrete pebbles 
more rapidly when $\alpha$ is small. Our results suggest that planets grow much 
more rapidly in disks with $\alpha \lesssim 10^{-4}$. 

Stochastic processes are another feature of the evolution.  In calculations with 
$\alpha = 10^{-4}$, it takes $2-5 \times 10^4$ yr to produce an ensemble of massive 
oligarchs \citep{kok98} with $M_p \gtrsim 3 \times 10^{26}$ g (Figure \ref{fig:noli}). 
For a large set of calculations with identical starting conditions, the random timing 
of oligarch formation and variations in accretion rates for each oligarch lead to 
a broad range in $N_{o,max}$, the maximum number of oligarchs. Often a rapidly accreting 
oligarch prevents a neighboring oligarch from accreting pebbles and reaching the 
promotion mass. Thus, calculations with small $N_{o,max}$ tend to have more massive
oligarchs than calculations with large $N_{o,max}$.

Mergers of protoplanets are the final important feature of the growth of gas giant planets 
\citep[see also][]{thom02,kb06,ida2010,li2010}. As protoplanets grow to masses of 5--10 \mearth,
their orbital separations are typically 10--20 Hill radii. Thus, mergers are rare.
Once protoplanets start to accrete gas, their separations rapidly become smaller
than 5--10 Hill radii. Strong dynamical interactions among pairs of protoplanets 
begin. Some interactions scatter protoplanets into the inner disk at 1--3 AU;
others scatter protoplanets into the outer disk at $\gtrsim$ 30--100 AU
\citep[see also][2009b, 2010]{mar02,ver09,ray09a}.  Throughout 
this chaotic growth phase, mergers rapidly deplete the number of growing protoplanets
(Figure \ref{fig:noli}). Once the number of protoplanets reaches $\sim$ 5--10,
planets have typical separations of $\gtrsim$ 10 Hill radii. Chaotic growth ends.

The timescale for chaotic growth varies from one calculation to the next 
(Figure \ref{fig:noli}). Sometimes, protoplanets are widely separated, grow
slowly, and reach chaotic growth at late times (indigo and blue curves in 
Figure \ref{fig:noli}). When several closely packed protoplanets begin to 
accrete gas at roughly the same time, their growth initiates an early phase 
of chaotic growth (magenta and turquoise curves in Figure \ref{fig:noli}). 
In nearly all cases, chaotic growth ceases on timescales of 3--10~Myr.

With growth rates sensitive to $\alpha$, the mass distributions of planets also depend 
on $\alpha$ (Figure \ref{fig:prob10}). In calculations with $M_{d,0}$ = 0.1 \msun\ and
$R_{d,0}$ = 30 AU, disks with small $\alpha$ produce a variety of gas giants, with
masses ranging from 10 \mearth\ to 10--20 \mjup. Disks with larger and larger $\alpha$ 
yield smaller and smaller gas giants. For all $\alpha$, the probability of a given mass 
is roughly a power law, $p \propto M_P^{-n}$, with $n$ = 0.25--1. Thus, these calculations
produce more Earth mass planets than Jupiter mass planets \citep[see also][]{ida05,mor2009}. 
The exponent in the power law depends on $\alpha$: models with smaller $\alpha$ produce 
broader probability distributions with smaller $n$.

For all $\alpha$, gas giants have a broad range of core masses\footnote{For simplicity,
we assume that all of the solid material in a planet is in the core \citep[see][]{hell2008}.}. 
Although gas accretion
begins when the core mass is 10 \mearth, planets continue to accrete solid pebbles.
During chaotic growth, protoplanets wander through a large range in orbital semimajor 
axis $a$, sweeping up (or scattering) pebbles and smaller oligarchs along their orbits.  Mergers 
of two gas giants also augment the core mass. For disks with $M_{d,0}$ = 0.1 \msun,
core masses range from 10 \mearth\ to 100 \mearth.

Our calculations suggest that Jupiter mass or larger planets form throughout the disk 
(Figure \ref{fig:jsemi}). For $\alpha \lesssim 10^{-4}$, Jupiters achieve fairly stable 
orbits at semimajor axes $a_J \approx$ 2--30~AU. Although most 10 \mearth\ cores form
at $a \sim$ 4--10~AU, dynamical interactions scatter protoplanets to $a \approx$ 1--3~AU 
and to $a \gtrsim$ 15--20~AU \citep[see also][]{tsi05}. After the scattering events, 
these planets continue to accrete material and can grow to gas giant planets in their 
new orbits.

Planets with smaller masses have broader distributions of semimajor axes than massive gas giants.
For convenience, we divide lower mass planets into `Saturns' with masses of 15 \mearth\ to
1 \mjup\ and `super-Earths' with masses of 1--15 \mearth. For large $\alpha \sim$
$10^{-3} - 10^{-2}$, Saturn-mass planets are often the most massive planet in the system
and have orbits with $a_S \approx$ 3--30 AU (Figure \ref{fig:nsemi}). When $\alpha$ is
small, $\approx 10^{-4} - 10^{-5}$, Saturns are scattered into orbits with $a_S \gtrsim$ 
100 AU, well outside those of Jupiter-mass planets. A few Saturn-mass planets are ejected
from the system. Super-Earths have even broader distributions of semimajor axis. All
super-Earths avoid regions close to gas giant planets (Figure \ref{fig:esemi}). Although 
a few super-Earths are scattered into the inner disk, most are scattered into the outer 
disk.  Many super-Earths are ejected.

Lower mass disks produce correspondingly lower mass planets.  Figure \ref{fig:prob04} shows 
predicted mass distributions for planets in a disk with $M_{d,0}$ = 0.04 \msun. Disks with
$\alpha \gtrsim 10^{-3}$ fail to produce Jupiter mass planets; disks with smaller $\alpha$
rarely produce 3--20 \mjup\ planets.  The semimajor axes of these planets are much more
concentrated towards the central star. The rare Jupiter mass planets lie within 10 AU; 
Saturn mass planets occupy 3--30 AU. Super-Earths occupy the largest range in $a$, with
$a_{SE} \approx$ 1--100 AU.

Our results suggest that lower mass disks fail to produce gas giants for any $\alpha$. In
disks with $M_{d,0}$ = 0.02~\msun\ (Figure \ref{fig:prob02}), disks with small $\alpha$
sometimes produce ice giants with $M_p \approx$ 15--30~\mearth. However, most planets in
these simulations are super-Earths with $M_p \approx$ 1--15~\mearth. Nearly all of these 
planets lie at small semimajor axes, $a_{SE} \approx$ 3--10~AU.

Finally, these calculations produce a diverse set of multi-planet systems (Table \ref{tbl: pfreq}).  
For any $M_{d,0}$, disks with smaller $\alpha$ yield a larger frequency of systems with at least 
two planets.  Defining $M_{p,max}$ as the maximum planet mass, planetary systems with smaller 
$M_{p,max}$ tend to have more planets. Systems with 4 Jupiter mass planets are rare: 5--10\%
among 0.1~\msun\ disks and less than 1\% among lower mass disks. Systems with 4--10 super-Earths
are relatively common, with frequencies of 70\% or more among 0.01--0.02 \msun\ disks.

\section{SUMMARY}

The current version of our hybrid code now includes all of the necessary physics 
to calculate the formation of gas giant planets from an input ensemble of planetesimals
in an evolving gaseous disk. The new features of the code include

\begin{enumerate}

\item the time evolution of a viscous disk with mass loss from photoevaporation and 
gas giant planet formation,

\item the atmospheric structure of planets with $R \approx$ 
0.001--10~\rearth\ with an algorithm to calculate the enhanced radius of a 
planet and the accretion of small dust grains,

\item gas accretion onto Earth-mass icy cores,

\item the time evolution of the radius and luminosity of gas giants, 

\item the gravitational potential of a massive gaseous disk and its impact on the orbits
of planets, and

\item the gravitational perturbation of (i) $N$-bodies on a disk of planetesimals and 
(ii) a disk of massive planetesimals on $N$-bodies.

\end{enumerate}

The new hybrid code accurately treats the migration of planets through a disk of
massive planetesimals. Simulations with isolated planets reproduce the results of 
\citet{kir09}, where the planet migrates inwards through undisturbed planetesimals.
Simulations with multiple planets reproduce the results of \citet{hah99}. These two 
examples illustrate the importance of the dynamical state of the planetesimals.  In 
the simulations for Fig.~\ref{fig:nep}, the initial environment of the planetesimals
orbiting near Neptune is similar to the initial conditions near the single planet in
the calculations for Fig.~\ref{fig:migrate_ae}--\ref{fig:migrate_ap}. Yet, the migration
rates in the two cases differ in magnitude and sign. Bromley \& Kenyon (2011) investigate
this phenomenon in more detail.

To begin to understand the orbital migration of ensembles of gas giants in an evolving,
gaseous disk, we consider a new `type 0' migration mechanism. Here, the semimajor axes
of planets evolve as the disk mass changes. Although the degree of migration is limited
by the ratio of the mass of the disk to the mass of the host star, modest inward or 
outward migration is possible. Because angular momentum is invariant, the sign of the
migration depends on the mode of mass and angular momentum loss. In systems that lose 
mass by disk accretion, migration is inward. Mass loss by photoevaporation may lead to
outward migration. Both modes may be important in establishing pairs of planets in 
resonant orbits \citep[e.g., as in][]{hol2010}.

To explore the ability of our code to simulate the formation of gas giant planets, we 
consider the evolution of planetesimals in disks with a variety of initial masses and 
viscosity parameters.  Disks of Pluto-mass planetesimals cannot form gas giants. For 
disks with initial masses of 0.1~\msun\ and all values of $\alpha$, the maximum planet 
mass is roughly 0.5~\mearth. Lower mass disks produce substantially lower mass planets.

Disks composed of 0.1--1~cm pebbles and a few Pluto-mass seeds can produce Jupiter mass
planets on short timescales.  Our results demonstrate that the properties of planetary 
systems depend on the initial disk mass $M_{d,0}$ and the viscosity parameter $\alpha$
(Figure \ref{fig:schema}).

\begin{enumerate}

\item More massive disks produce more massive planets on more rapid timescales. In disks
with $M_{d,0}$ = 0.1 \msun, Jupiter-mass planets are common at 1--2~Myr. In lower mass
disks, super-Earths and Saturns form in 2--3~Myr.  The timescales are similar to the 
observed lifetimes, $\sim$ 1--3 Myr, of circumstellar disks around low mass pre-main 
sequence stars \citep[e.g.,][]{hai01,kenn09,mamajek2009}.

\item Disks with larger $\alpha$ form lower mass planets. In disks with $\alpha = 10^{-2} - 10^{-3}$,
Jupiters form rarely; Saturns and Super-Earths are common. In disks with $\alpha = 10^{-4} - 10^{-5}$, 
planetary systems with 2 or more Jupiters are common. This result suggests that Jupiters
probably form more often in disks with `dead zones,' regions where the disk viscosity 
parameter is much lower than the rest of the disk \citep[e.g.,][]{mat09}.

\item The derived distributions of planet mass spans the range of known exoplanets. For
planets with $M_P \approx$ 0.03--10~\mjup, disks with $M_{d,0}$ = 0.1 \msun\ and $\alpha$
= $10^{-5} - 10^{-4}$ yield a roughly power law probability distribution
$dp / d~{\rm log}~M_p$ $\propto M_P^{-n}$ with $n$ = 0.20--0.25. Disks with $\alpha$
= $10^{-2}$ have a much steeper relation, with $n \approx$ 0.8--0.9.  Known exoplanets have
an intermediate frequency distribution, with $n \approx$ 0.48 \citep{how2010}. Thus, 
simulations with small (large) $\alpha$ produce relatively fewer (more) super-Earths and 
Saturns than observed in real systems. If real protostellar disks have a range of $\alpha$
that includes our small and large $\alpha$ regimes, some mix of $\alpha$ can probably
`match' the observations. However, including other physical processes -- such as migration
and photoevaporation -- will change the slopes of the mass-frequency relation 
\citep[see, for example,][]{ida05,mor2009}. Thus, detailed comparison with observations is 
premature.

\item The derived distribution of semimajor axes is roughly linear in $a$ over 3--30 AU.
Scattering leads to a few gas giants at 1--3 AU and at 30--100 AU.  Without a treatment 
of migration, our predicted distribution of semimajor axes cannot hope to match observations. 
However, our ability to scatter gas giants out to 30--100 AU may allow core accretion models
to produce systems similar to HR 8799 \citep{mar2008}.

\item These calculations often produce planetary systems with 2--10 planets. Planetary systems
with multiple planets
usually contain super-Earths, sometimes contain Saturns, and rarely contain Jupiters.  Current 
observations suggest that roughly 10\% of all planetary systems have two or more gas giant 
and/or ice giant planets \citep[][also statistics at exoplanet.eu and exoplanets.org]{gou2010}. 
However, some planetary systems may have as many as 7 super-Earth or ice giant planets 
\citep[e.g., HD10180;][]{lov2010}. Given current poor constraints on initial disk mass and
viscosity in protostellar disks, our results can `match' both of these observations.  

\item Jupiter mass planets in these calculations have core masses of 10--100 \mearth. Once
icy 10~\mearth\ cores start to accrete gas, they continue to accrete solids from the disk.
Mergers with other gas giants also augment the core mass. When mergers between growing
protoplanets are rare, planets have modest core masses, 10--20 \mearth.  When mergers between
icy cores are common, core masses approach 100 \mearth. Multiple mergers of growing 
protoplanets may account for the large heavy element abundances in HD 149026b and other 
exoplanets \citep{burr2007, fort2009}.

\end{enumerate}

These results are encouraging.  Our simple set of initial calculations accounts for the
observed range of planet masses and for some of the observed range of semimajor axes. The
derived power law slopes for the frequency of planet masses bracket the observations. 
Compared with real planetary systems, the calculations probably produce too many 
multi-planet systems. 

Adding more realism (e.g., migration, photoevaporation, etc) to our simulations will allow 
us to improve our understanding of the processes that lead to ice/gas giant planet formation.
Integrating orbits past $\sim$ 10~Myr will enable robust comparisons between the properties
of real and simulated planetary systems. Together with improved observations of protoplanetary
disks and better data for exoplanets, numerical simulations like ours should lead to clear
tests for the core accretion theory of planet formation.

\acknowledgements
We acknowledge generous allotments of computer time on the NASA `discover' cluster 
($\sim$ 1000 cpu days) and on the SI cluster `hydra' ($\sim$ 500 cpu days). Advice and 
comments from A. Burrows, T. Currie, M. Geller, G. Kennedy, G. Marcy, R. Murray-Clay,
H. Perets, D. Spiegel, and A. Youdin greatly improved our presentation.  We thank an
ananymous referee for comments that clarified many parts of our discussion.  Portions 
of this project were supported by the {\it NASA} 
{\it Astrophysics Theory} and {\it Origins of Solar Systems} programs 
through grant NNX10AF35G, the {\it NASA} {\it TPF Foundation Science Program,} through 
grant NNG06GH25G, and the {\it Spitzer Guest Observer Program,} through grant 20132.

\appendix

\section{The gravitational potential of an unperturbed disk}\label{appx:disk}

If the gaseous protostellar disk is extended, thin, and axisymmetric, our code can treat 
the gravitational effect of the disk on the planets.  We estimate the gravitational 
potential produced by a geometrically thin, axisymmetric disk in two ways. The first 
method is based on an analytical inversion of the Poisson equation from an expansion of 
the surface density $\Sigma$ into Bessel functions. The results are simple expressions 
for the potential, valid for 2D disks of large spatial extent and a surface density that 
is a pure power-law in orbital distance.  The second method employs brute force, using 
a direct numerical integration of the mass density over a geometrically thin, axisymmetric 
3D disk. This approach gives the potential associated with an arbitrary surface density 
profile, with the advantage of generalizability at the expense of computational load.

The analytical approach for estimating the disk potential as it acts
on a unit-mass particle takes advantage of solutions to the Laplace
equation that are linear combinations of Bessel functions of the first
kind, $J_0(x)$ \citep[\S 2.6]{bt87}
\begin{equation}\label{eq:Phiaz}
\Phid(R,z) = \int_0^\infty S(k) J_0(kR) e^{-k|z|} dk,
\end{equation}
where $S(k)$ is a density function. The exponential with the
discontinuous first derivative at $z = 0$ generates a $\delta$
function in the 3D density at the disk midplane upon application of
the Laplace operator. Gauss's law relates this potential to the
surface density. A closed, pillbox-shaped surface with a unit-area
cross-section that straddles the disk midplane has a gravitational
field flux $-2 d\Phid/dz$, and encloses a mass $\Sigma$. Gauss's law
sets the flux equal to $4\pi G$ times the enclosed mass, hence
\begin{equation}\label{eq:SigmaSk}
\Sigma(R) = -\frac{1}{2\pi G} \int_0^\infty S(k) J_0(kR) k dk .
\end{equation}
The integral is a Hankel transform (to within a multiplicative
constant), and we identify $S(k)$ and $\Sigma(R)$ as transform pairs,
with
\begin{equation}\label{eq:Hankel}
S(k) = -2\pi G \int_0^\infty \Sigma(R) J_0(kR) R dR .
\end{equation}

We now assume that the surface density $\Sigma(R)$ is a power
law in $R$ with an arbitrarily large radial extent, i.e., $\rin
\rightarrow 0$ and $\rout \rightarrow \infty$. In this case,
\begin{equation}\label{eq:Sigmapower}
\Sigma(a) = \Sigma_0 (R_0/R)^{-\powlaw}
\end{equation}
and
\begin{equation}\label{eq:Sk}
S(k) = -2\pi G \Sigma_0 R_0^{\powlaw} k^{\powlaw -2} 
      \int_0^\infty q^{1-\powlaw} J_0(q) dq \ .
\end{equation}
We impose the condition that $1/2 < \powlaw < 2$ so that the mass in the
disk is finite inside a finite radius, with the lower limit set  so that the
integral in equation~\ref{eq:Sk} is well-defined. This integral is of 
the form
\begin{eqnarray}\label{eq:hankel}
I_n(\mu)  & = & \int_0^\infty x^\mu J_n(x) dx  \\
      \ & = & 2^\mu 
      \frac{\powlaw(\frac{n+1+\mu}{2})}{\powlaw(\frac{n+1-\mu}{2})} \\
      \ & = & \frac{1}{I_n(-\mu)} \ ,
\end{eqnarray}

\begin{equation}
S(k) \approx -2\pi G 
   \Sigma_0 R_0^{\powlaw} k^{\powlaw -2} I_0(2-\powlaw) \ .
\end{equation}
Similarly, the potential in the disk midplane is
\begin{equation}
\Phid(R) = -2\pi G\Sigma_0 
R \left(\frac{R}{R_0}\right)^{-\powlaw} I_0(1-\powlaw) I_0(\powlaw-2) ,
\end{equation}
although this expression is formally valid only for $1 < \powlaw <
2$. The radial component of the
acceleration, $a_r$, is not subject to this formal restriction on $\powlaw$;
it obtains from the derivative of $\Phid$ with respect
to $R$, along with the identity $dJ_0(x)/dx = -J_1(x)$:
\begin{equation}
a_r = -2\pi G \Sigma_0 
I_0(1-\powlaw) I_1(1-\powlaw) 
\left(\frac{R}{a_0}\right)^{-\powlaw},
\end{equation}
which is exact in the midplane of an infinitely extended disk.

To illustrate results from the above analysis, we give the
following examples of the potential and radial acceleration for a few
power-law surface density profiles:
\begin{equation}
\begin{array}{lll}
\powlaw = 0.6: & \Phid = 2.8\pi G\Sigma_0 R_0^{0.6}R^{0.4} & \ \ \ \
a_r = -1.1\pi G\Sigma_0 (R/R_0)^{-0.6}
\medskip\\ 
\powlaw = 1.0 & \Phid = 2\pi G \Sigma_0 R_0 \log(R) + {\rm const} &  \ \ \ \
a_r = -2\pi G \Sigma_0 R_0/R  
\medskip\\
\powlaw = 1.5 & \Phid = -8.8\pi G\Sigma_0 R_0^{1.5}R^{-0.5} &  \ \ \ \
a_r = -4.4\pi G \Sigma_0 (R/R_0)^{-1.5}
\end{array}
\end{equation}
As long as the orbital distance $R$ approximately satisfies $\rin < 0.1 R$ and $10 R
< \rout$, these expressions hold for disks of finite extent.

As an alternative to the analytical method, which is
limited in terms of the functional form of $\Sigma(R)$, we take a
numerical approach to directly integrate the mass in the disk to get
the gravitational potential:
\begin{equation}
\label{eq:Phi}
\Phid(\vec{r},t) = \int_\rin^\rout \int_{-\pi}^{\pi} \frac{dR d\varphi R
  \Sigma(R,t)}
{\sqrt(r_\perp^2 + z^2 + h^2 +R^2 - 2 r_\perp R   
\cos(\varphi)} ,
\end{equation}
where position $\vec{r}$ relative to the host star is in cylindrical
coordinates, $(r_\perp,\varphi,z)$. The variable $h$ is a softening
parameter, which formally eliminates the density singularity in the
disk midplane by making the 2D disk behave as if it has some finite
thickness.  From the Poisson equation, we find that the effective
vertical density profile is approximately Gaussian, with a scale
height of $\sigma \sim h$. Here we take $h = 0.01$~AU; in what
follows, the results do not strongly depend on this choice.

The next step is to integrate eq.~(\ref{eq:Phi}) numerically; 
the radial integration can be solved analytically, but integrating
over the angular coordinate $\varphi$ requires a Chebyshev-Legendre 
quadrature scheme. The number of sample points $n$ in our quadrature 
scheme is variable; we increase $n$ by a factor of four successively 
until the relative change in the result from one iteration to the next 
is below an error tolerance of $10^{-7}$. For the integrands encountered 
here, the result is much more accurate than the error tolerance suggests, 
because the quadrature scheme typically converges quadratically or better.

Compared to with evaluating the $1/r$ potential from the central star,
this method is computationally expensive. To alleviate this problem, we 
limit the calculations to the disk midplane---a reasonable approximation 
since the orbital inclination of planets not entrained in the gas is 
typically less than the disk scale height.  Then we evaluate the potential 
at a set of logarithmically spaced points in orbital separation and 
interpolate with cubic splines. The computational load then reduces to 
$O(10)$ arithmetic operations.

\section{Time-varying disk mass (``type 0'' migration)}\label{appx:type0}

When the disk mass is removed from the planetary system by photoevaporation
or some other slow process, the time-varying disk mass can produce a radial
migration of a planet's orbit.  Angular momentum is conserved.  For example, 
the product of the semimajor axis $a$ and the central star mass $M$ remains 
constant when the star itself loses mass. For low-eccentricity orbits, the 
conserved quantity is equal to $a^3 a_r$, where $a_r$ is the acceleration on 
the body from both the star and gaseous disk. If the disk mass---but not the
stellar mass---varies in time, 
\begin{eqnarray}
a(t) & = & a(0)\frac{1 + 2\pi \Sigma_0/M I_0(1-\powlaw) I_1(1-\powlaw) 
                     a_0^{\powlaw} a(0)^{2-\powlaw}}{1+M
2\pi \Sigma_0/M I_0(1-\powlaw) I_1(1-\powlaw) a_0^\powlaw 
                     a(t)^{2-\powlaw} e^{-t/\tau}} 
\\
\nonumber
\ & \ & \
\\
\ & \approx & \
\left[1 + 2\pi \Sigma_0/M I_0(1-\powlaw) I_1(1-\powlaw) 
a_0^{\powlaw} a(0)^{2-\powlaw}\right] \ \ \ \ (t \gg \tau) \, .
\end{eqnarray}
Setting $\powlaw = 1$ and taking the limit of $t \rightarrow \infty$,
the maximum change in orbital distance for a planet at 1~AU\ in an 
evaporating disk with initial surface density $\Sigma0 = 3000$~g/cm$^2$ 
is roughly 1\%.  If a disk with the same mass is more centrally concentrated 
($\powlaw = 1.5$),  this number is roughly an order of magnitude higher.
For planets farther out in the disk, the maximum change is much larger,
$\sim$ 5\% at 10 AU and $\sim$ 10\% at 30 AU.

\vfill
\eject

\clearpage
\begin{deluxetable}{lcccccccccc}
\tablecolumns{11}
\tablewidth{0pc}
\tabletypesize{\small}
\tablecaption{Frequency of Multi-Planet Systems\tablenotemark{1}}
\tablehead{
  \colhead{} &
  \colhead{} &
  \multicolumn{9}{c}{Number of Planets}
\\
  \colhead{$\alpha$} &
  \colhead{Mass Key} &
  \colhead{0} &
  \colhead{1} &
  \colhead{2} &
  \colhead{3} &
  \colhead{4} &
  \colhead{5} &
  \colhead{6} &
  \colhead{7} &
  \colhead{8}
}
\startdata
\cutinhead{$M_{d,0}$ = 0.1 \msun}
$10^{-2}$ & S & 0.30 & 0.16 & 0.30 & 0.16 & 0.05 & 0.00 & 0.00 & 0.00 & 0.00 \\
$10^{-3}$ & J & 0.30 & 0.35 & 0.35 & 0.00 & 0.00 & 0.00 & 0.00 & 0.00 & 0.00 \\
$10^{-4}$ & J & 0.00 & 0.00 & 0.19 & 0.70 & 0.11 & 0.00 & 0.00 & 0.00 & 0.00 \\
$10^{-5}$ & J & 0.00 & 0.02 & 0.25 & 0.68 & 0.05 & 0.00 & 0.00 & 0.00 & 0.00 \\
\cutinhead{$M_{d,0}$ = 0.04 \msun}
$10^{-2}$ & SE & 0.00 & 0.00 & 0.33 & 0.37 & 0.28 & 0.02 & 0.00 & 0.00 & 0.00 \\
$10^{-3}$ & S & 0.47 & 0.37 & 0.16 & 0.00 & 0.00 & 0.00 & 0.00 & 0.00 & 0.00 \\
$10^{-4}$ & J & 0.18 & 0.65 & 0.17 & 0.00 & 0.00 & 0.00 & 0.00 & 0.00 & 0.00 \\
$10^{-5}$ & J & 0.06 & 0.55 & 0.35 & 0.04 & 0.00 & 0.00 & 0.00 & 0.00 & 0.00 \\
\cutinhead{$M_{d,0}$ = 0.02 \msun}
$10^{-2}$ & SE & 0.97 & 0.03 & 0.00 & 0.00 & 0.00 & 0.00 & 0.00 & 0.00 & 0.00 \\
$10^{-3}$ & SE & 0.00 & 0.00 & 0.02 & 0.21 & 0.33 & 0.37 & 0.07 & 0.00 & 0.00 \\
$10^{-4}$ & S & 0.97 & 0.03 & 0.00 & 0.00 & 0.00 & 0.00 & 0.00 & 0.00 & 0.00 \\
$10^{-5}$ & S & 0.91 & 0.09 & 0.00 & 0.00 & 0.00 & 0.00 & 0.00 & 0.00 & 0.00 \\
\cutinhead{$M_{d,0}$ = 0.01 \msun}
$10^{-2}$ & SE & 1.00 & 0.00 & 0.00 & 0.00 & 0.00 & 0.00 & 0.00 & 0.00 & 0.00 \\
$10^{-3}$ & SE & 0.98 & 0.02 & 0.00 & 0.00 & 0.00 & 0.00 & 0.00 & 0.00 & 0.00 \\
$10^{-4}$ & SE & 0.00 & 0.00 & 0.00 & 0.31 & 0.28 & 0.28 & 0.13 & 0.00 & 0.00 \\
$10^{-5}$ & SE & 0.00 & 0.00 & 0.00 & 0.00 & 0.00 & 0.12 & 0.31 & 0.29 & 0.28 \\
\enddata
\tablenotetext{1}{Fraction of planets in systems where the most massive planet 
is a Jupiter ('J' in Mass Key column), a Saturn ('S'), or a super-Earth ('SE').
For example, in the first row, disks with $M_{d,0}$ = 0.1 \msun\ and $\alpha = 
10^{-2}$ produce Saturn or lower mass planets; the frequency of systems with $N$
Saturn-mass planets is 30\% ($N$ = 0), 16\% ($N$ = 1), 30\% ($N$ = 2), 16\% ($N$ = 3), 
5\% ($N$ = 4) and 0\% ($N \ge$ 5).}
\label{tbl: pfreq}
\end{deluxetable}
\clearpage

\clearpage

\clearpage

\clearpage

\clearpage

\clearpage

\clearpage

\begin{figure}
\includegraphics[width=6.0in]{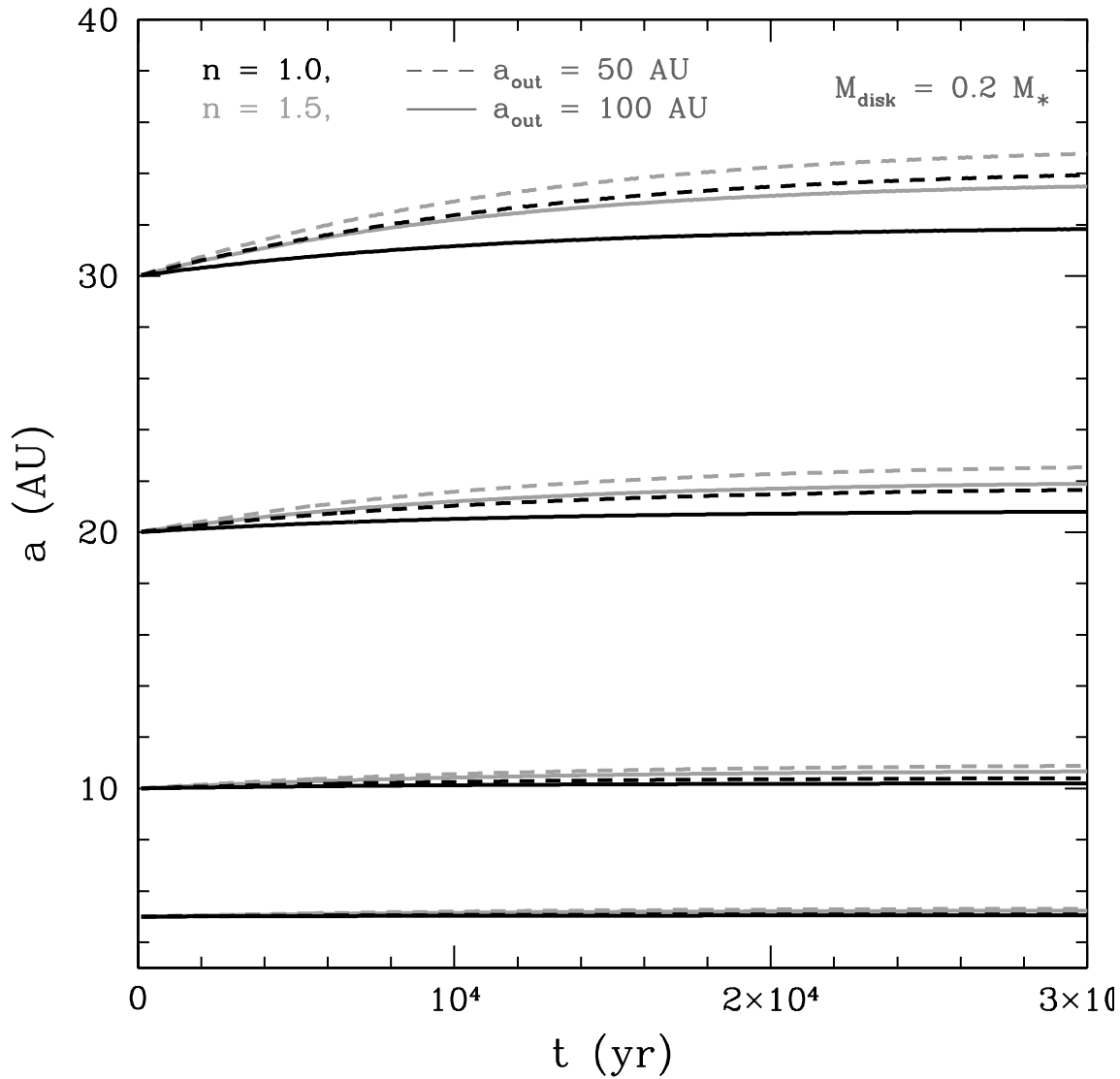}
\caption{Semimajor axis as a function of time in Type 0
migration. Planets move in response to a decrease in disk mass, for
four different massive disks, with surface density $\Sigma \sim
a^{-\powlaw}$, and outer disk radius $a_{out}$ as labeled. The initial
mass of each disk is 20\% of the stellar mass (1~\msun); the disk mass 
decays exponentially on a timescale of $10^4$~yr. The greatest migration 
is for the $n = 1.5$ disk with an inner edge of 50~AU (gray dashed line), 
which has the most mass initially concentrated within the orbits of the
planets.
\label{fig:type0}
}
\end{figure}
\clearpage

\begin{figure}
\includegraphics[width=6.5in]{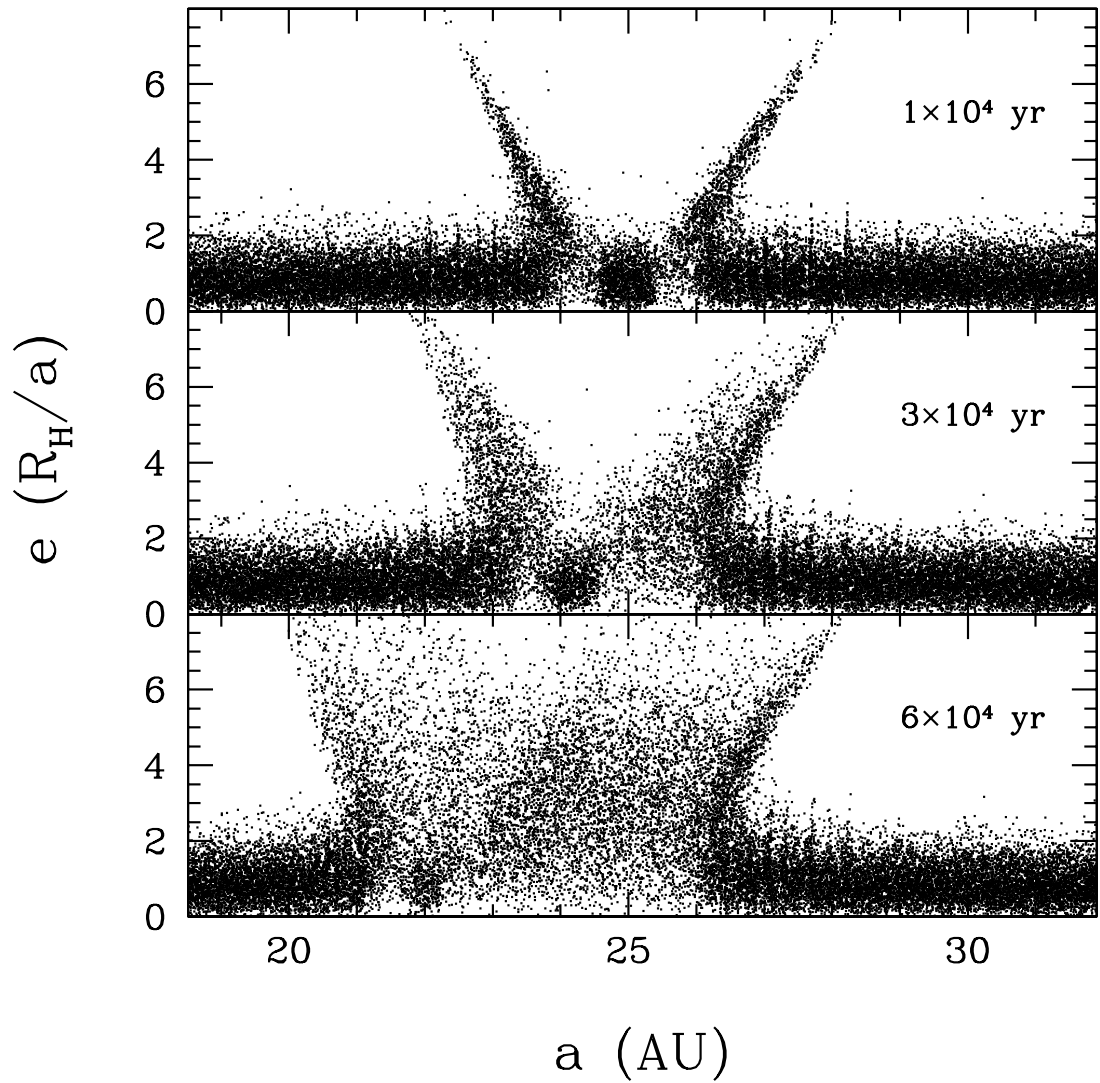}
\caption{The evolution of eccentricity in units of Hill radii as a
function of semimajor axis during planet migration. The three panels
show snapshots at the indicated times as a 2~\mearth\ planet travels
through a sea of planetesimals, similar to \citet{kir09} --- $\Sigma
\sim 1/a$, normalized to 1.2~g~cm$^{-2}$ at 25~AU, and extending from
14.5~AU to 35.5~AU. The blob of particles in each panel ($\sim$ 25 AU 
at $10^4$ yr, $\sim$ 24 AU at $3 \times 10^4$ yr, and $\sim$ 22 AU at
$6 \times 10^4$ yr) indicates the location of the planet.
\label{fig:migrate_ae}
}
\end{figure}
\clearpage

\begin{figure}
\includegraphics[width=6.5in]{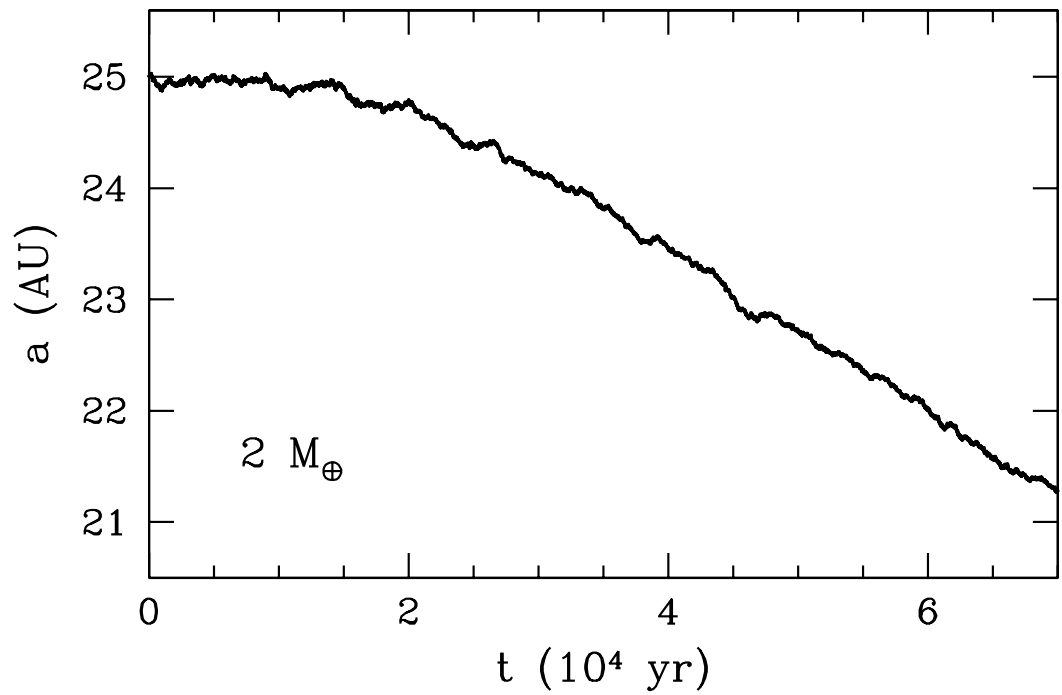}
\caption{The evolution of the semimajor axis of a planet in the
planetesimal disk from Figure~\ref{fig:migrate_ae}. The initially
flat migration profile stems from ``inertia'' associated with
planetesimals in the co-orbital zone (the blobs of particles in
Fig.~\ref{fig:migrate_ae}, which diffuse away in time). After the
planet scatters these particles out of its orbit, fast migration 
begins.
\label{fig:migrate_ap}
}
\end{figure}
\clearpage

\begin{figure}
\includegraphics[width=6.5in]{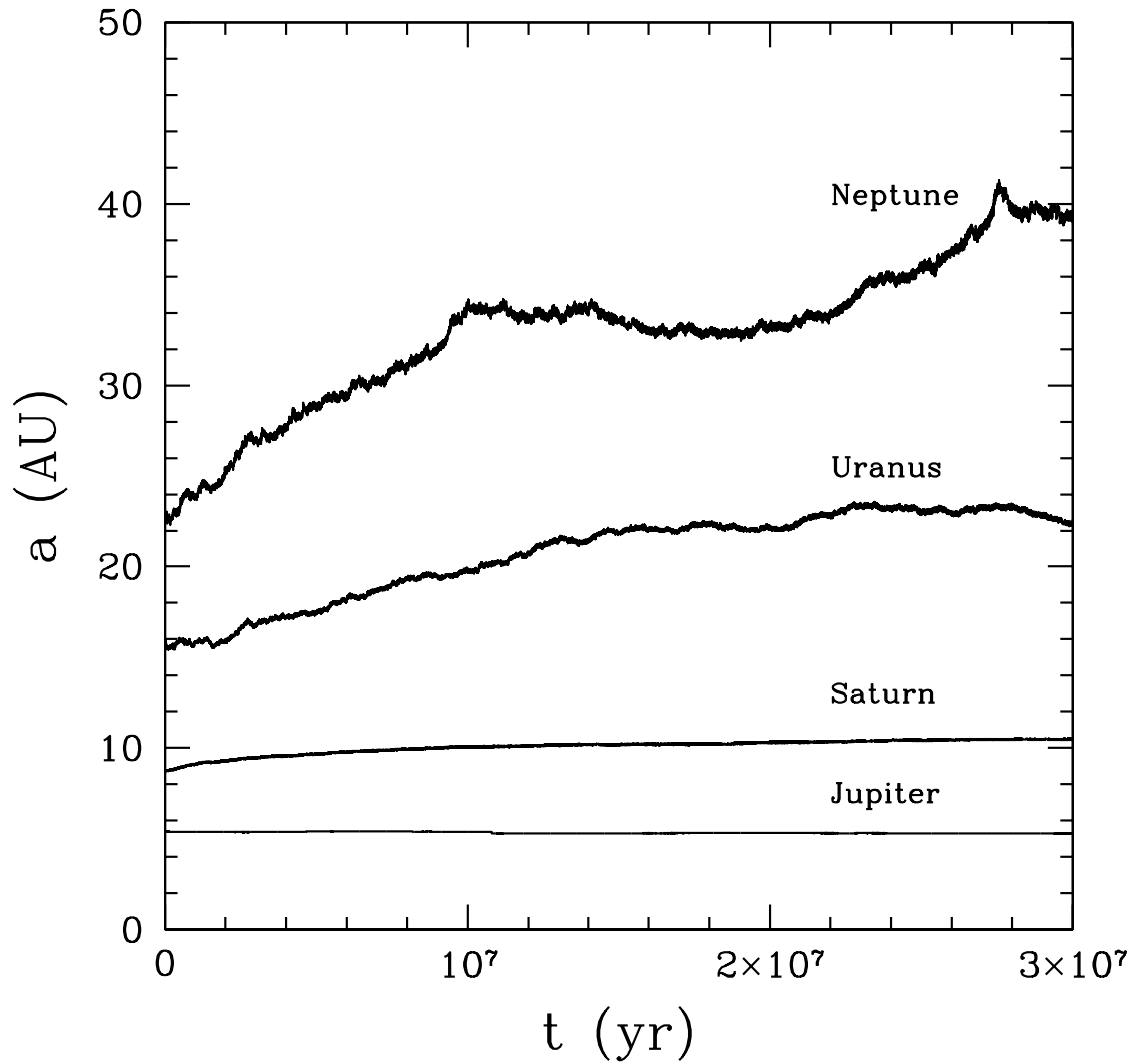}
\vskip -3ex
\caption{Migration of the major planets in a planetesimal disk. The
planets, with their present-day masses, evolve from an initially
compact configuration as a result of interactions with 1000
planetesimals, each with a mass of 0.1~\mearth, uniformly distributed
in semimajor axis from 10~AU to 50~AU.  In contrast to the behavior of 
the single planet in Fig.~\ref{fig:migrate_ap}, interactions between
planetesimals and multiple planets cause the outward migration of Neptune 
and Uranus \citep[see][]{hah99}.
\label{fig:nep}
}
\end{figure}
\clearpage

\begin{figure}
\includegraphics[width=6.5in]{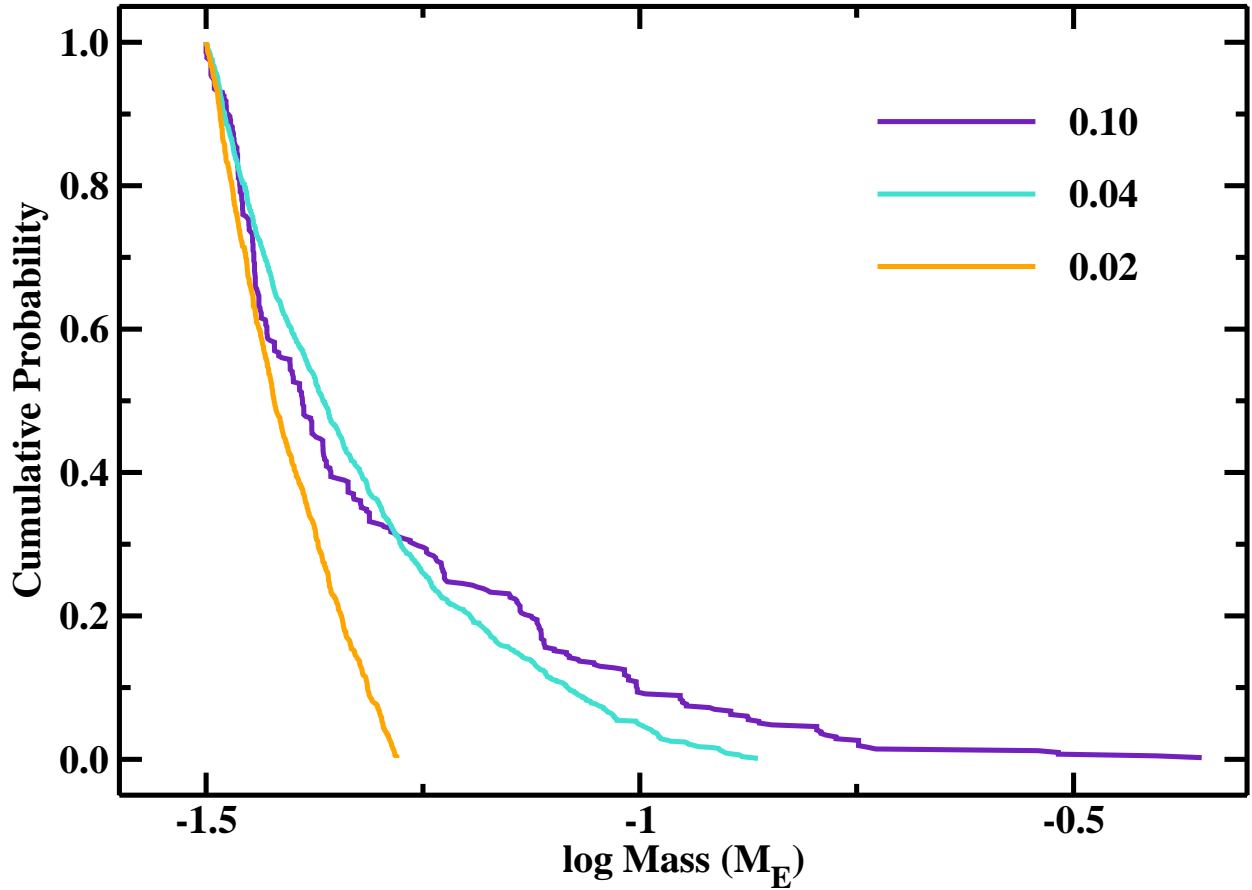}
\vskip 3ex
\caption{Mass distributions at 3 Myr for calculations with ensembles of
Pluto-mass planetesimals. The legend indicates the initial disk
mass in \msun.  The maximum planet mass scales with the disk mass.
For planets with masses exceeding 10 \mearth, the cumulative 
probability is roughly a power law, $p \propto M_P^{-\beta}$. 
Less massive disks have steeper probability distributions.
Disks with initial masses $M_{d.0} \lesssim $ 0.01 \msun\ do not
produce 10 \mearth\ planets.
\label{fig:pluto-prob}
}
\end{figure}
\clearpage

\begin{figure}
\includegraphics[width=6.5in]{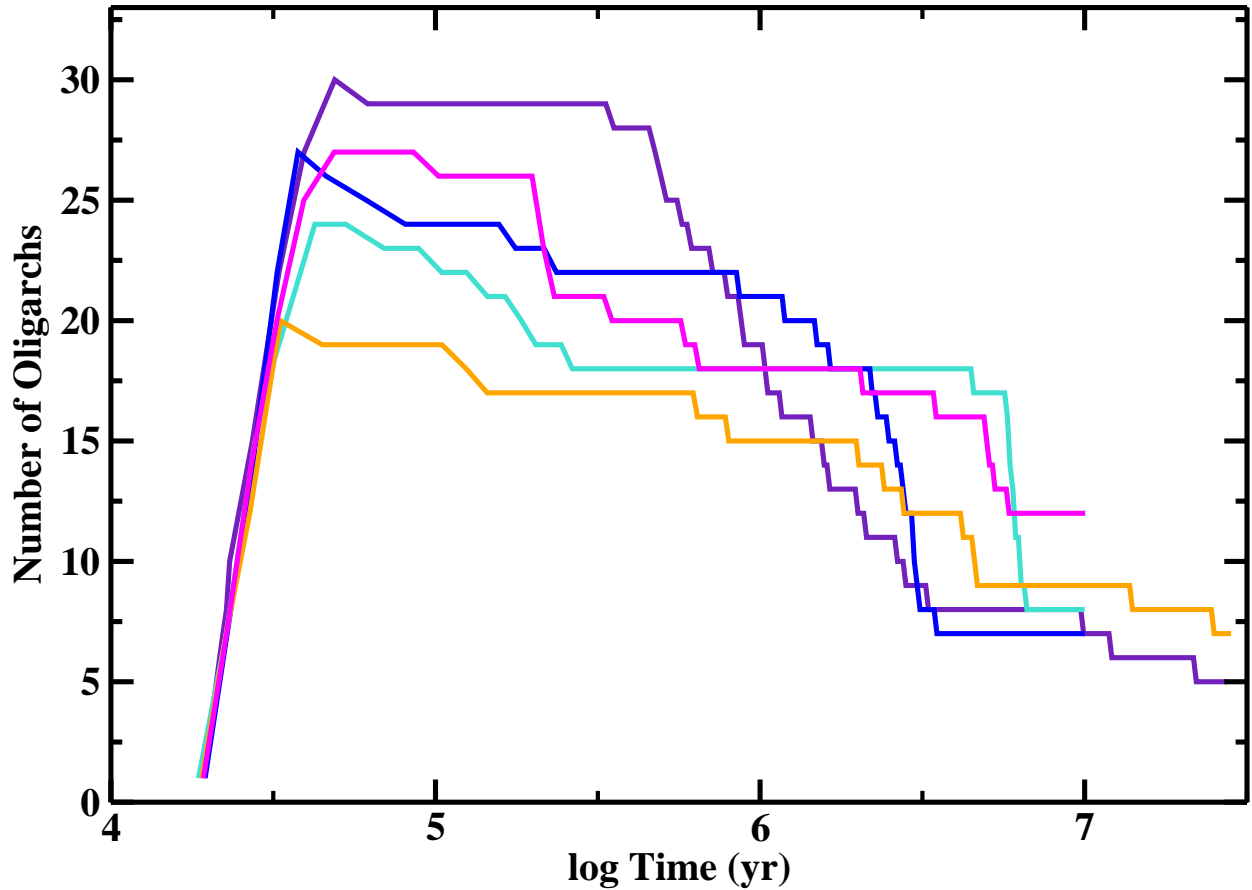}
\vskip 3ex
\caption{Time evolution of $N_{oli}$, the number of oligarchs with 
$M_p > 3 \times 10^{26}$, in models with $M_{d,0}$ = 0.1 \msun,
$R_{d,0}$ = 30 AU, and $\alpha = 10^{-3}$. In these simulations,
$N_{oli}$ rises as Pluto mass seeds rapidly accrete tiny pebbles,
reaches a plateau when these seeds begin to accrete gas, and then
falls as gas giant planets merge with other gas giants and smaller
oligarchs. As gas giants achieve stable orbits, $N_{oli}$ stabilizes.
\label{fig:noli}
}
\end{figure}
\clearpage

\begin{figure}
\includegraphics[width=6.5in]{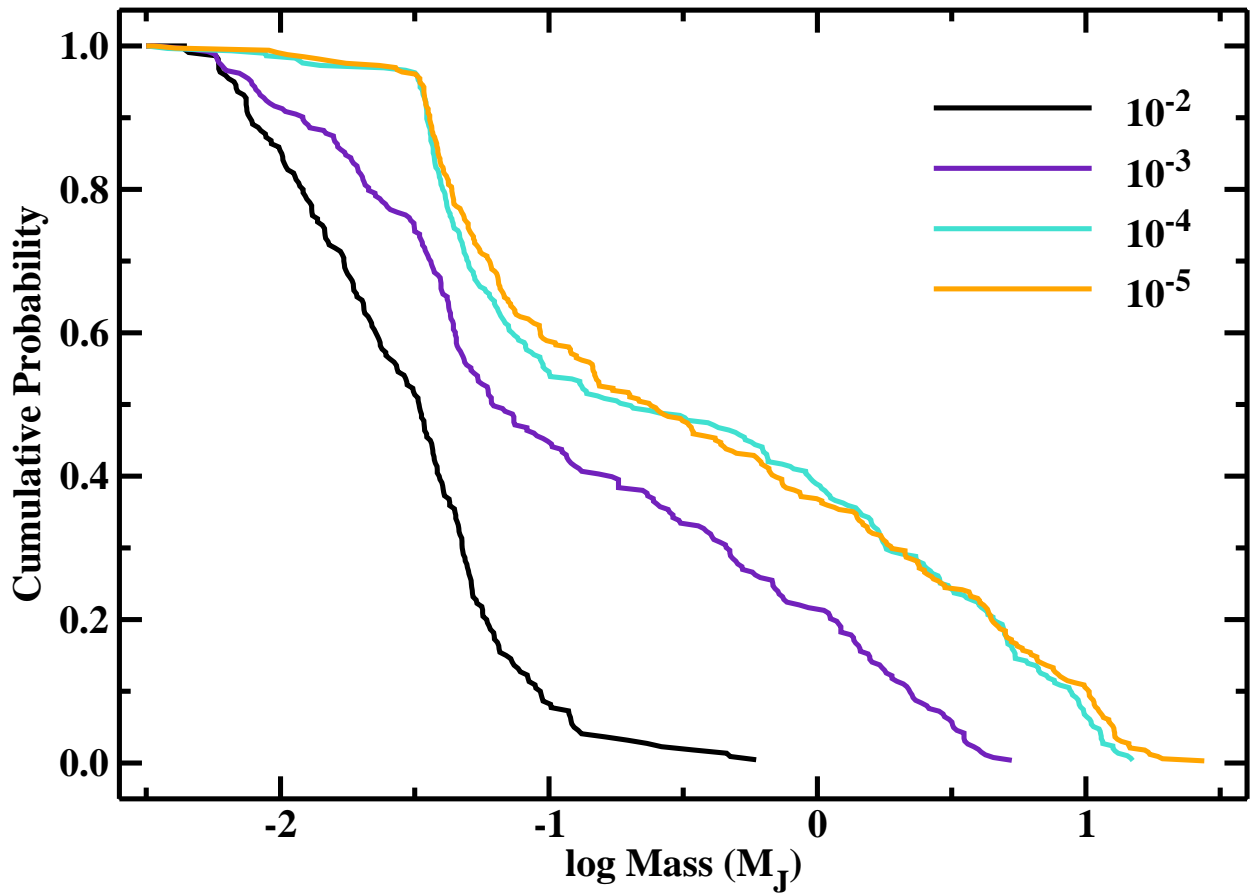}
\vskip 3ex
\caption{Mass distributions at 3 Myr for disks with $M_{d,0}$ = 0.1 \msun,
$R_{d,0}$ = 30 AU, and $\alpha$ as indicated in the legend. Disks
with smaller $\alpha$ produce more massive gas giants. The median mass 
of gas giants ranges from $\sim$ 20 \mearth\ for $\alpha = 10^{-2}$
to $\sim$ 0.25 \mjup\ for $\alpha = 10^{-5}$.
\label{fig:prob10}
}
\end{figure}
\clearpage

\begin{figure}
\includegraphics[width=6.5in]{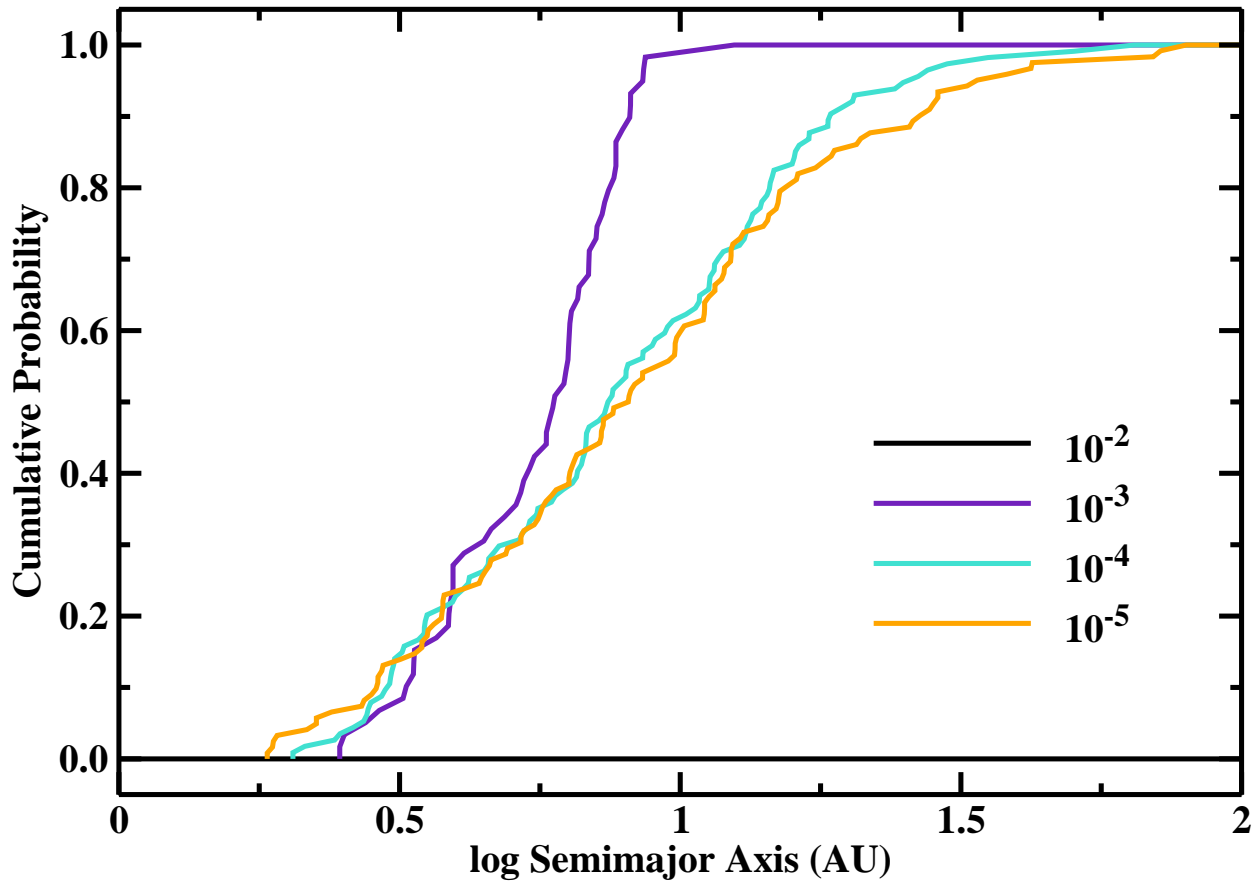}
\vskip 3ex
\caption{As in Fig. \ref{fig:prob10} for the semimajor axes of planets
with masses larger than 1~\mjup. Disks with smaller $\alpha$ produce 
gas giants over a broader range of semimajor axes. The median semimajor 
axis ranges from $\approx$ 6~AU for $\alpha = 10^{-3}$ to $\approx$ 8 AU 
for $\alpha = 10^{-5}$. Disks with $\alpha = 10^{-2}$ do not make massive
gas giant planets.
\label{fig:jsemi}
}
\end{figure}
\clearpage

\begin{figure}
\includegraphics[width=6.5in]{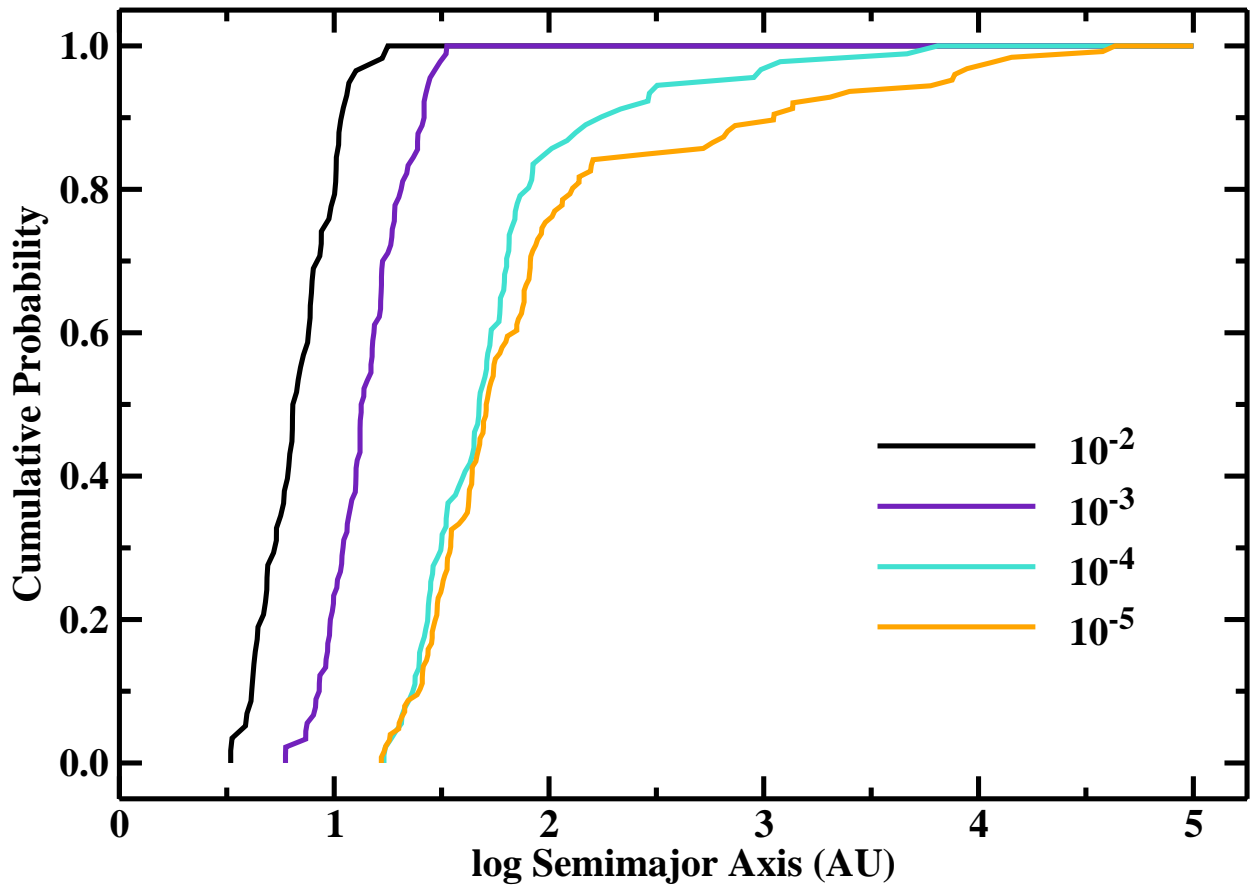}
\vskip 3ex
\caption{As in Fig. \ref{fig:jsemi} for the semimajor axes of planets
with masses between 15~\mearth\ and 1~\mjup. Disks with smaller $\alpha$
produce Neptune to Jupiter mass planets at larger semimajor axes.  
At 3 Myr, the median semimajor axis ranges from $\approx$ 6~AU for 
$\alpha = 10^{-2}$ to $\approx$ 50 AU for $\alpha = 10^{-5}$.
\label{fig:nsemi}
}
\end{figure}
\clearpage

\begin{figure}
\includegraphics[width=6.5in]{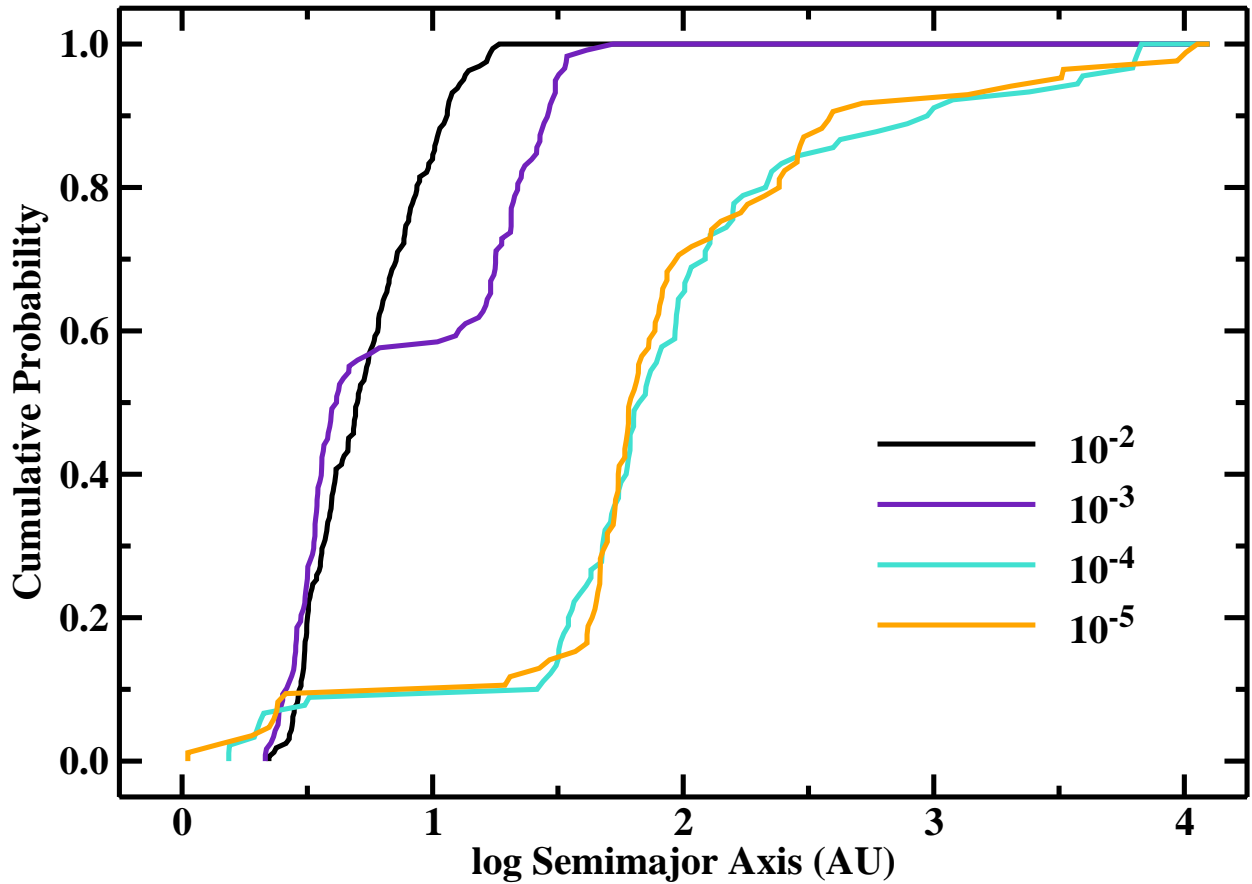}
\vskip 3ex
\caption{As in Fig. \ref{fig:jsemi} for the semimajor axes of planets
with masses 1--15 \mearth. Disks with smaller $\alpha$ produce super-Earth
mass planets over a broader range of semimajor axes.  Super-Earths avoid
regions near Jupiter mass planets.  At 3 Myr, the median semimajor axis 
ranges from $\approx$ 5~AU for $\alpha = 10^{-2}$ to $\approx$ 60 AU for 
$\alpha = 10^{-5}$.
\label{fig:esemi}
}
\end{figure}
\clearpage

\begin{figure}
\includegraphics[width=6.5in]{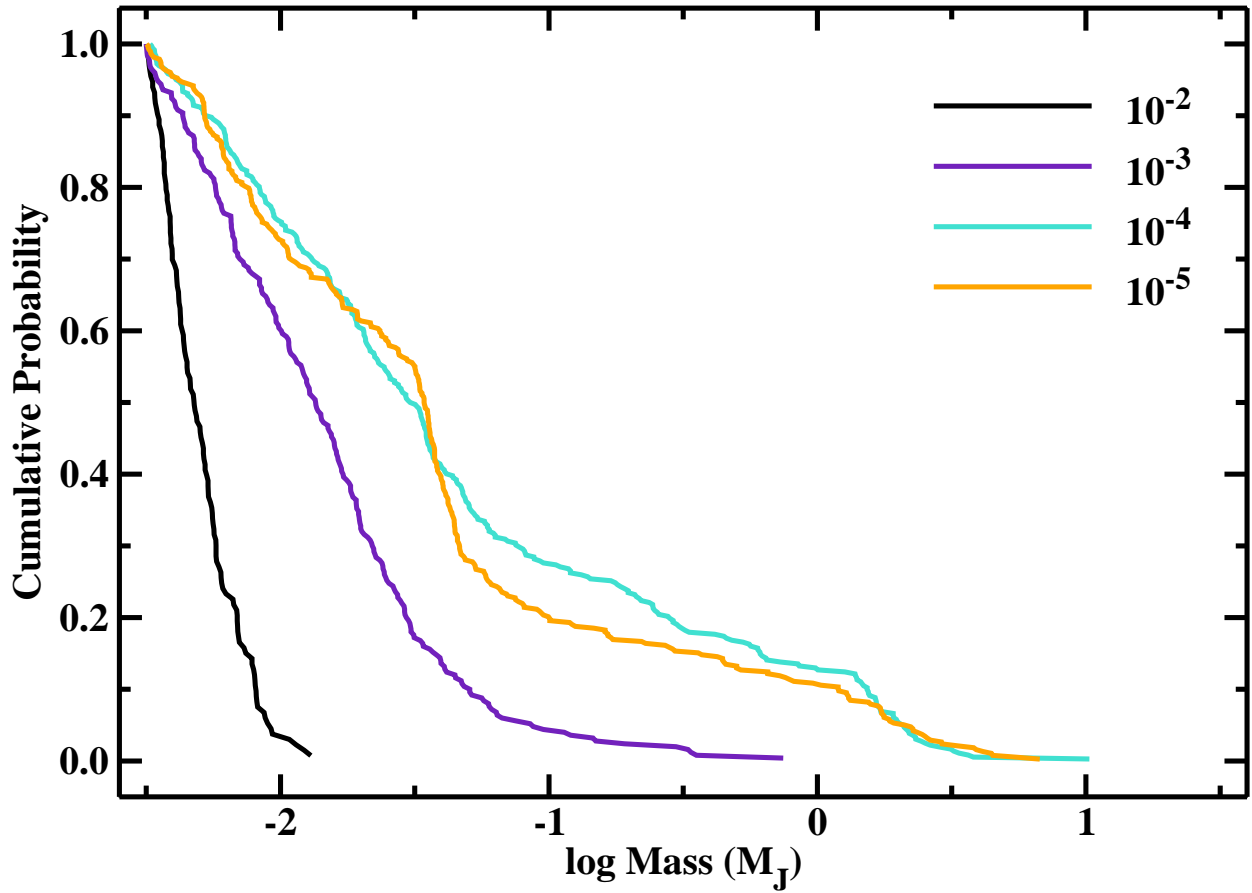}
\vskip 3ex
\caption{Mass distributions at 3 Myr for disks with $M_{d,0}$ = 0.04 \msun,
$R_{d,0}$ = 30 AU, and $\alpha$ as indicated in the legend. Disks
with smaller $\alpha$ produce more massive gas giants. The median mass 
of planets ranges from $\sim$ 1.5 \mearth\ for $\alpha = 10^{-2}$
to $\sim$ 20 \mearth\ for $\alpha = 10^{-5}$.
\label{fig:prob04}
}
\end{figure}
\clearpage

\begin{figure}
\includegraphics[width=6.5in]{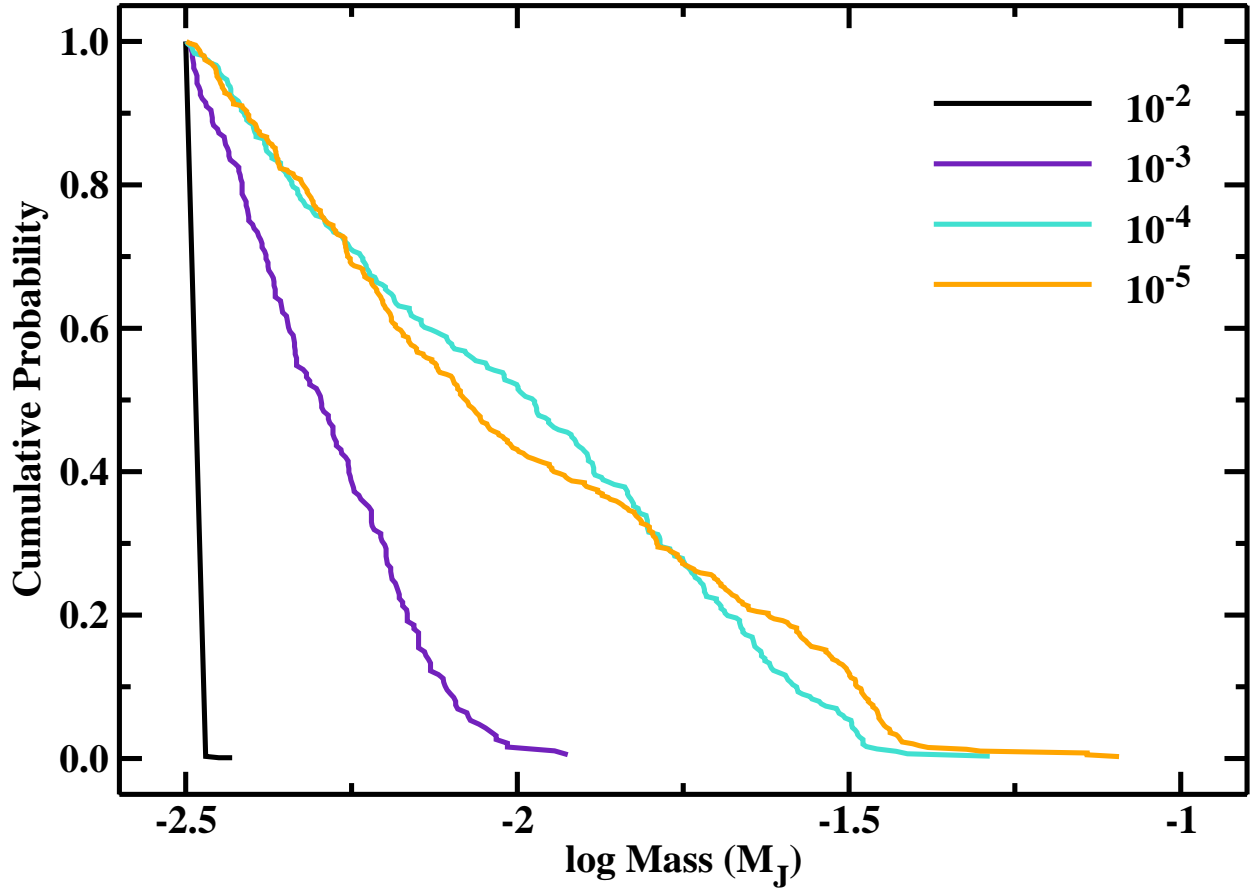}
\vskip 3ex
\caption{As in Figure \ref{fig:prob04} for disks with $M_{d,0}$ = 0.02 \msun.
Note the change of scale for the abscissa compared to other figures.  The 
median mass ranges from $\sim$ 1 \mearth\ for $\alpha = 10^{-2}$ to $\sim$ 
6 \mearth\ for $\alpha = 10^{-5}$.
\label{fig:prob02}
}
\end{figure}
\clearpage

\begin{figure}
\includegraphics[width=6.5in]{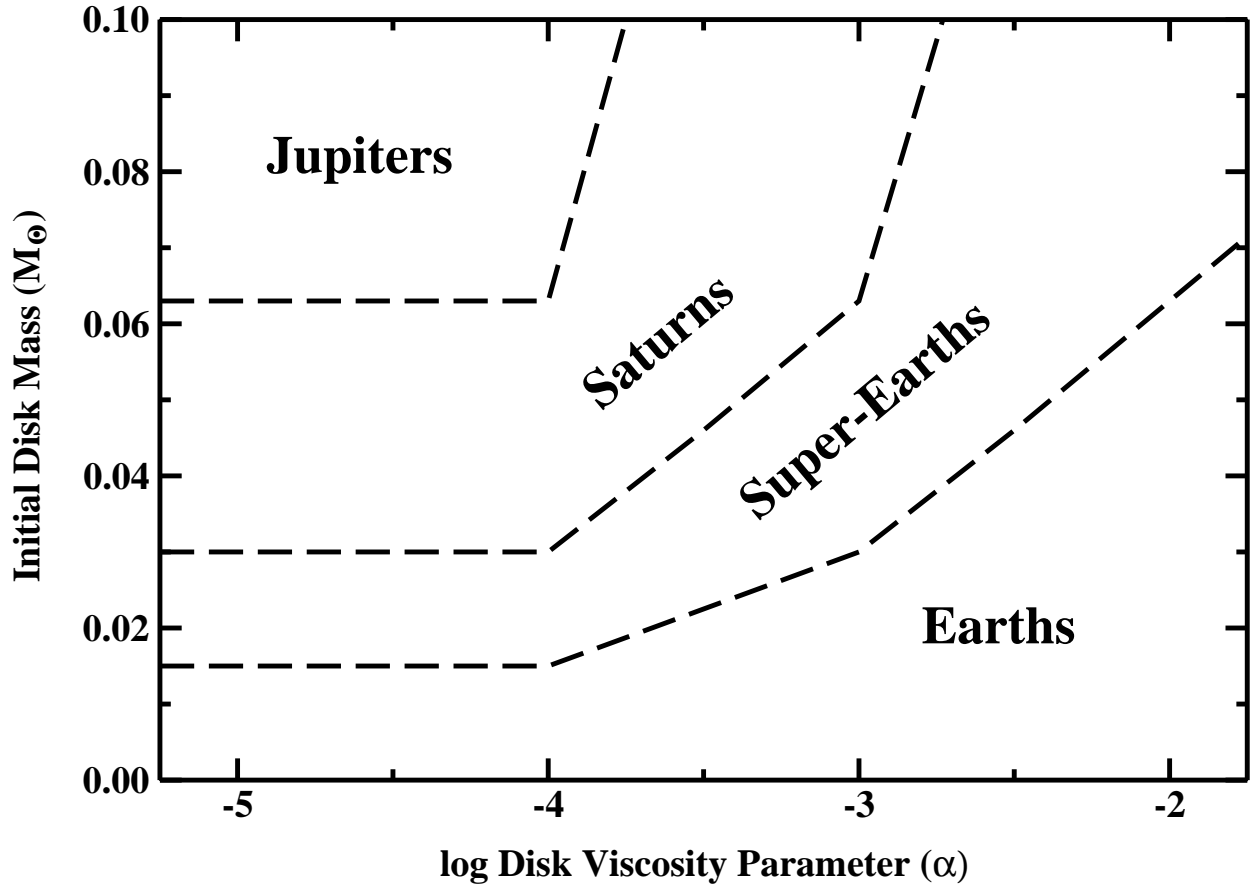}
\vskip 3ex
\caption{
Outcomes of planet formation simulations.
The regions indicate the maximum masses of planets as
a function of the initial disk mass, $M_{d,0}$, and
the disk viscosity parameter, $\alpha$.
\label{fig:schema}
}
\end{figure}
\clearpage

\end{document}